\documentclass[aps,prd,twocolumn,nofootinbib,showpacs,longbibliography]{revtex4-1}
\usepackage{bm,amsmath,amssymb,graphicx}
\usepackage[colorlinks,linkcolor=blue,urlcolor=blue]{hyperref}
\begin{document}

\title{Mathematical structure and physical content of composite gravity in weak-field approximation}

\author{Hans Christian \"Ottinger}
\email[]{hco@mat.ethz.ch}
\homepage[]{www.polyphys.mat.ethz.ch}
\affiliation{ETH Z\"urich, Department of Materials, CH-8093 Z\"urich, Switzerland}

\date{\today}

\begin{abstract}
The natural constraints for the weak-field approximation to composite gravity, which is obtained by expressing the gauge vector fields of the Yang-Mills theory based on the Lorentz group in terms of tetrad variables and their derivatives, are analyzed in detail within a canonical Hamiltonian approach. Although this higher derivative theory involves a large number of fields, only few degrees of freedom are left, which are recognized as selected stable solutions of the underlying Yang-Mills theory. The constraint structure suggests a consistent double coupling of matter to both Yang-Mills and tetrad fields, which results in a selection among the solutions of the Yang-Mills theory in the presence of properly chosen conserved currents. Scalar and tensorial coupling mechanisms are proposed, where the latter mechanism essentially reproduces linearized general relativity. In the weak-field approximation, geodesic particle motion in static isotropic gravitational fields is found for both coupling mechanisms. An important issue is the proper Lorentz covariant criterion for choosing a background Minkowski system for the composite theory of gravity.
\end{abstract}


% 04.20.-q   Classical general relativity
% 04.20.Cv   Fundamental problems and general formalism
% 04.50.Kd   Modified theories of gravity

\maketitle

\section{Introduction}\label{secintroduction}
Einstein's general theory of relativity may not be the final word on gravity. As beautiful and successful as it is, it seems to have serious problems both on very small and on very large length scales. A problem on small length scales is signaled by $90$ years of unwavering resistance of general relativity to quantization. A problem on the largest length scales is indicated by the present search for ``dark energy'' to explain the accelerated expansion of the universe within general relativity. These problems provide the main motivation for continued research on alternative theories of gravity (see, for example, the broad review \cite{CapozzielloDeLau11} of extended theories of gravity).

A composite higher derivative theory of gravity has recently been proposed in \cite{hco231}. The general idea of a composite theory is to specify the variables of a ``workhorse theory'' in terms of more fundamental variables and their time derivatives \cite{hco235,hco237}. The occurrence of time derivatives in the ``composition rule'' leads to a higher derivative theory, which is naturally tamed by the constraints resulting from the composition rule. For the composite theory of gravity proposed in \cite{hco231}, the underlying workhorse theory is the Yang-Mills theory \cite{YangMills54} based on the Lorentz group and the composition rule expresses the corresponding gauge vector fields in terms of the \emph{tetrad} or \emph{vierbein} variables providing a Cholesky-type decomposition of a metric. As a consequence of the composite structure of the proposed theory, it differs significantly from contentious previous attempts \cite{Utiyama56,Yang74,BlagojevicHehl} to turn the Yang-Mills theory based on the Lorentz group into a theory of gravity.

Whereas the original formulation of composite gravity in \cite{hco231} was based on the Lagrangian framework, we here switch to the Hamiltonian approach. As a Hamiltonian formulation separates time from space, it certainly cannot provide the most elegant formulation of relativistic theories. However, the Hamiltonian framework has clear advantages by offering a natural formulation of constraints and a straightforward canonical quantization procedure. For bringing constraints and quantization together, we here establish the constraints resulting from the composition rule as second class constraints that can be treated via Dirac brackets \cite{Dirac50,Dirac58a,Dirac58b}, whereas gauge constraints can be handled separately by BRST quantization (the acronym derives from the names of the authors of the original papers \cite{BecchiRouetStora76,Tyutin75}; see also \cite{Nemeschanskyetal86,hco229}). Moreover, the Hamiltonian approach provides the natural starting point for a generalization to dissipative systems. In particular, this approach allows us to formulate quantum master equations \cite{BreuerPetru,Weiss,hco199,hco221} and to make gravity accessible to the robust framework of dissipative quantum field theory \cite{hcoqft}.

The number of fields involved in the composite theory of gravity is enormously large. Each Yang-Mills vector field has  four components satisfying  second-order differential equations so that, in the Hamiltonian approach, four additional conjugate momenta are required. For the Lorentz group, the six Yang-Mills vector fields associated with six infinitesimal generators (three rotations, three boosts) thus result in $6 \times 8 = 48$ fields. Gauge constraints eventually reduce this number of degrees of freedom by a factor of two (simply speaking, among the four components of a vector field, only the two transverse components carry physical information). In addition, we consider $16$ tetrad or vierbein variables, again coming with conjugate momenta, so that we deal with a total of $48+32=80$ fields in our canonical Hamiltonian approach. Actually, this is not even the end of the story as additional ghost fields would be introduced in the BRST approach for handling the gauge constraints. Our approach differs from the traditional Hamiltonian formulation of general higher derivative theories developed by Ostrogradsky \cite{Ostrogradsky1850,Woodard15,Gitmanetal83}. The Ostrogradsky framework would involve only $4 \times 16 = 64$ fields, but would possess much less structure and less natural constraints \cite{hco235}. A key task of the present paper is to elaborate in detail in the context of the linearized theory that the constraints from the composition rule, together with the gauge constraints, reduce this enormous number of fields to just a few degrees of freedom, as expected for a theory of gravity. Another important task of the present discussion of the weak-field approximation is to provide guidance for the discussion of the fully nonlinear composite theory of gravity. Understanding the structure of the constraints is helpful also for proper coupling of the gravitational field to matter. Whereas the coupling of the Yang-Mills fields to matter was considered previously \cite{hco231}, we here introduce a properly matched additional coupling of the tetrad fields to matter. 

The structure of the paper is as follows. In a first step, we introduce the space of $80$ fields for our canonical Hamiltonian formulation of linearized composite gravity, with special emphasis on gauge transformations and the implications of the composition rule (Section~\ref{secarena}). For the pure field theory in the absence of matter, we elaborate all evolution equations and constraints in detail and we readily find the solutions for gravitational waves and static isotropic systems (Section~\ref{secpurefields}). We subsequently introduce a double coupling mechanism for Yang-Mills and tetrad fields to matter into composite gravity. The modifications resulting from the inclusion of matter are elaborated to obtain a complete theory of gravity that can be compared to linearized general relativity (Section~\ref{secwithmatter}). We finally summarize our results and draw a number of conclusions (Section~\ref{secconclusions}). The relation between the Lagrangian and Hamiltonian approaches and some intermediate and additional results are provided in three appendices.

\section{Arena for composite theory}\label{secarena}
For developing the composite theory of gravity, we consider a fixed background Minkowski space where $x^0=ct$ is the product of the speed of light and time, $x^1,x^2,x^3$ are the spatial coordinates, and $\eta_{\mu\nu} = \eta^{\mu\nu}$ denotes the Minkowski metric [with signature $(-,+,+,+)$]. Greek indices go from $0$ to $3$. The Minkowski metric, which is its own inverse, is always used for raising or lowering space-time indices. Throughout this paper we set the speed of light equal to unity ($c=1$). Assuming a background Minkowski space comes with the advantage of offering a clear understanding of energy, momentum and their conservation laws.

\subsection{Tetrad variables and gauge vector fields}
Standard \emph{tetrad} or \emph{vierbein} variables ${b^\kappa}_\mu$ result from a Cholesky-type decomposition of a metric $ g_{\mu\nu} $,
\begin{equation}\label{Choleskyg}
   g_{\mu\nu} = \eta_{\kappa\lambda} \, {b^\kappa}_\mu {b^\lambda}_\nu ,
\end{equation}
which may also be interpreted as a coordinate transformation associated with a local set of base vectors. The non-uniqueness of this decomposition is the source of the gauge transformation behavior discussed in the next subsection. In the weak-field approximation, we write
\begin{equation}\label{glinb}
   {b^\kappa}_\mu = {\delta^\kappa}_\mu + \eta^{\kappa\lambda} \hat{h}_{\lambda\mu}  ,
\end{equation}
where $\hat{h}_{\lambda\mu}$ is assumed to be small so that we need to keep only the lowest-order terms. It is convenient to define the symmetric and antisymmetric parts of $\hat{h}_{\mu\nu}$,
\begin{equation}\label{hhatsyms}
   h_{\mu\nu} = \hat{h}_{\mu\nu} + \hat{h}_{\nu\mu} , \qquad
   \omega_{\mu\nu} = \hat{h}_{\mu\nu} - \hat{h}_{\nu\mu} .
\end{equation}
In the weak-field approximation, we obtain the following first-order expression for the metric (\ref{Choleskyg}),
\begin{equation}\label{gling}
   g_{\mu\nu} = \eta_{\mu\nu} + h_{\mu\nu} .
\end{equation}

We denote the conjugate momenta associated with ${b^\kappa}_\mu$ by ${p_\kappa}^\mu$. Again, it is useful to introduce the symmetric and antisymmetric parts,
\begin{equation}\label{psyms}
   \tilde{h}^{\mu\nu} = p^{\mu\nu} + p^{\nu\mu} , \qquad
   \tilde{\omega}^{\mu\nu} = p^{\mu\nu} - p^{\nu\mu} .
\end{equation}
The $32$ fields ${b^\kappa}_\mu$ and ${p_\kappa}^\mu$ represent the canonical space associated with the tetrad variables, which in turn characterize a metric. The fields $\tilde{h}^{\mu\nu}$ and $\tilde{\omega}^{\mu\nu}$ may essentially be regarded as the conjugate momenta associated with $h_{\mu\nu}$ and $\omega_{\mu\nu}$, respectively (after properly accounting for normalization and symmetrization effects).

The Hamiltonian description of a Yang-Mills theory is based on the four-vector fields $A_{a \mu}$ and their conjugates $E^{a \mu}$, which are the generalizations of the vector potentials and the electric fields of electrodynamics, respectively. Whereas $\mu$ is the usual space-time index, $a$ labels the base vectors of the Lie algebra associated with the underlying Lie group. For the Lorentz group, which consists of the real $4 \times 4$ matrices that leave the Minkowski metric invariant, the Lie algebra is six-dimensional. We here choose six natural base vectors of the Lie algebra, three of which generate the Lorentz boosts in the coordinate directions and the other three generate rotations around the coordinate axes. It is convenient to switch back and forth between the labels $a=1, \ldots 6$ for all six generators and the pairs $(0,1)$, $(0,2)$, $(0,3)$ for the boosts in the respective directions (involving also time) and $(2,3)$, $(3,1)$, $(1,2)$ for the rotations in the respective planes according to Table~\ref{tabindexmatch}. In particular, we can now write our base vectors of the Lie algebra as
\begin{equation}\label{Lorentzgenerators}
   T^a_{\kappa\lambda} = {\delta^{\tilde{\kappa}}}_\lambda \, {\delta^{\tilde{\lambda}}}_\kappa -
   {\delta^{\tilde{\kappa}}}_\kappa \, {\delta^{\tilde{\lambda}}}_\lambda .
\end{equation}

\begin{table}
\begin{tabular}{c|c c c c c c}
	% \hline
    $a$ \, & \, $1$ & $2$ & $3$ & $4$ & $5$ & $6$ \\
	\hline
	$(\tilde{\kappa},\tilde{\lambda})$ \,
    & \, $(0,1)$ & $(0,2)$ & $(0,3)$ & $(2,3)$ & $(3,1)$ & $(1,2)$ \\
	% \hline
\end{tabular}
\caption{Correspondence between label $a$ for the base vectors of the six-dimensional Lie algebra ${\rm so}(1,3)$ and ordered pairs $(\tilde{\kappa},\tilde{\lambda})$ of space-time indices.}
\label{tabindexmatch}
\end{table}

We finally need to specify the composition rule for expressing the four-vector fields $A_{a \mu}$ in terms of the tetrad fields ${b^\kappa}_\mu$ or, in view of Eq.~(\ref{glinb}) equivalently, the symmetric and antisymmetric parts $h_{\mu\nu}$ and $\omega_{\mu\nu}$, respectively, of $\hat{h}_{\mu\nu}$. For $a=(\tilde{\kappa},\tilde{\lambda})$ according to Table~\ref{tabindexmatch}, we postulate the simple composition rule
\begin{equation}\label{glinA}
   A_{a \mu} = \frac{1}{2} \left(
   \frac{\partial h_{\tilde{\lambda}\mu}}{\partial x^{\tilde{\kappa}}} -
   \frac{\partial h_{\tilde{\kappa}\mu}}{\partial x^{\tilde{\lambda}}} \right)
   + \frac{1}{2\tilde{g}} \,
   \frac{\partial \omega_{\tilde{\kappa}\tilde{\lambda}}}{\partial x^\mu} ,
\end{equation}
where $\tilde{g}$ is a dimensionless coupling constant that controls the relative weight of the symmetric and antisymmetric contributions to ${b^\kappa}_\mu$. Only for $\tilde{g}=1$, the four-vector variables (\ref{glinA}) can be interpreted as a connection field \cite{hco231}. Such an interpretation would be essential for a closer relation to general relativity.

\subsection{Gauge transformations}
As the Minkowski metric in the decomposition (\ref{Choleskyg}) is invariant under Lorentz transformations, the corresponding transformation matrices can be applied to the factors ${b^\kappa}_\mu$ without changing the metric. For infinitesimal Lorentz transformations, this implies the lowest-order gauge transformation
\begin{equation}\label{gaugeb}
   \delta b_{\kappa\lambda} = - \tilde{g} \, \Lambda_a T^a_{\kappa\lambda} ,
\end{equation}
in terms of six additional fields $\Lambda_a$. As the base vectors $T^a_{\kappa\lambda}$ of the Lie algebra defined in Eq.~(\ref{Lorentzgenerators}) are antisymmetric, only the antisymmetric part of $b_{\kappa\lambda}$ is affected by gauge transformations, that is,
\begin{equation}\label{gaugeh}
   \delta h_{\kappa\lambda} = 0 ,
\end{equation}
whereas
\begin{equation}\label{gaugeom}
   \delta \omega_{\kappa\lambda} = - 2 \tilde{g} \, \Lambda_a T^a_{\kappa\lambda} .
\end{equation}
The latter equation suggests that the six fields $\Lambda_a$ can be chosen to make the six components of $\omega_{\kappa\lambda}$ equal to zero. We refer to this particular choice as the symmetric gauge.

From Eqs.~(\ref{glinA}), (\ref{gaugeh}) and (\ref{gaugeom}), we further obtain
\begin{equation}\label{gaugeA}
   \delta A_{a \mu} = \frac{\partial \Lambda_a}{\partial x^\mu} ,
\end{equation}
which is the proper gauge transformation behavior for the gauge-vector fields of the linearized theory. This transformation rule implies gauge invariance of the combination
\begin{equation}\label{gaugeinvcomb1}
   2 \tilde{g} A_{(\tilde{\kappa}\tilde{\lambda}) \mu}
   - \frac{\partial \omega_{\tilde{\kappa}\tilde{\lambda}}}{\partial x^\mu} ,
\end{equation}
which is obvious from the definition (\ref{glinA}) and the gauge invariance of $h_{\kappa\lambda}$.

We moreover assume that all conjugate momenta are gauge invariant (cf.\ Eq.~(49) of \cite{hco229}),
\begin{equation}\label{gaugeE}
   \delta E^{a \mu} = 0 ,
\end{equation}
and
\begin{equation}\label{gaugep}
   \delta p^{\kappa\lambda} = 0 .
\end{equation}
It turns out below that the assumption (\ref{gaugeE}) requires that $\partial A_{a \mu}/\partial x_\mu$ must be a gauge invariant quantity. In view of Eq.~(\ref{gaugeA}), this implies
\begin{equation}\label{Lamevoleq}
   \frac{\partial^2 \Lambda_a}{\partial x_\mu \partial x^\mu} = 0 .
\end{equation}
In other words, the fields $\Lambda_a$ generating gauge transformations must become dynamic players and satisfy free field equations. This idea is the basis of the BRST approach for handling gauge constraints. Moreover, we conclude
\begin{equation}\label{tetradom}
   \frac{\partial^2 \omega_{\kappa\lambda}}{\partial x_\mu \partial x^\mu} = 0 ,
\end{equation}
which follows from the symmetric gauge $\omega_{\kappa\lambda}=0$ and the gauge invariance of the left-hand side.

\subsection{Implications of composition rule}
The composition rule (\ref{glinA}) contains two types of equations. If $\tilde{\kappa}$ or $\mu$ is equal to zero, it contains time derivatives and hence implies time evolution equations for the tetrad variables. Otherwise, the composition rule provides constraints that must be satisfied at any time.

The expressions for $A_{(0k)l}+A_{(0l)k}$ and $A_{(kl)0}$ lead to the unambiguous evolution equations
\begin{eqnarray}
   \frac{\partial h_{kl}}{\partial t} &=& \frac{1}{2} \left( \frac{\partial h_{0l}}{\partial x^k}
   + \frac{\partial h_{0k}}{\partial x^l} \right) \nonumber \\
   &+&  A_{(0k)l} + A_{(0l)k}
   - \frac{1}{2\tilde{g}} \left( \frac{\partial \omega_{0l}}{\partial x^k}
   + \frac{\partial \omega_{0k}}{\partial x^l} \right)  , \qquad 
\label{hklevol}
\end{eqnarray}
and
\begin{equation}\label{omklevol}
   \frac{\partial \omega_{kl}}{\partial t} = 2 \tilde{g} A_{(kl)0}
   - \tilde{g} \left( \frac{\partial h_{0l}}{\partial x^k} -
   \frac{\partial h_{0k}}{\partial x^l} \right) .
\end{equation}
The expression for $A_{(0l)0}$ provides only the time derivative of $\tilde{g} h_{0l} + \omega_{0l}$, and there is no evolution equation for $h_{00}$ whatsoever. Once we have made a decision about the evolution of $h_{0\mu}$, all evolution equations are fixed uniquely.

Choosing four conditions for $h_{0\mu}$ is superficially reminiscent of imposing coordinate conditions for obtaining unique solutions in general relativity, but the logical status is entirely different. Whereas the coordinate conditions of general relativity have no influence on the physical predictions of general relativity, in the canonical formulation of composite gravity suitable conditions for $h_{0\mu}$ are used to characterize ``good'' or ``valid'' Minkowskian coordinate systems. If these conditions are Lorentz covariant, we have no possibility of switching between different types of conditions corresponding to different physical predictions. As obvious as these remarks may be, the proper appreciation of coordinate conditions in general relativity was a slow process, in which even Einstein could not easily detach himself from the idea of physically preferred coordinate systems \cite{Giovanelli20}.

An appealing set of Lorentz covariant conditions is given by
\begin{equation}\label{harmoniclin}
   \frac{\partial h_{\mu\nu}}{\partial x_\nu} =
   K \, \frac{\partial {h^\nu}_\nu}{\partial x^\mu} ,
\end{equation}
where $K=1/2$ corresponds to particularly convenient harmonic coordinates (in the linear approximation). We here adopt the conditions (\ref{harmoniclin}) as the tentative criteria for physically meaningful coordinates. They can be rewritten as explicit time evolution equations, namely
\begin{equation}\label{h0kevol}
   \frac{\partial  h_{0l}}{\partial t} = \frac{\partial h_{ln}}{\partial x_n}
   - K \, \frac{\partial {h^\nu}_\nu}{\partial x^l} ,
\end{equation}
and
\begin{equation}\label{h00evol}
   \frac{\partial  h_{00}}{\partial t} = \frac{\partial h_{0l}}{\partial x_l} 
   - \frac{K}{1-K} \left[ 2 A_{(0l)l}
   - \frac{1}{\tilde{g}} \frac{\partial \omega_{0l}}{\partial x_l} \right] .
\end{equation}
From the expression for $A_{(0l)0}$, we finally obtain
\begin{equation}\label{om0kevol}
   \frac{\partial \omega_{0l}}{\partial t} = 2 \tilde{g} A_{(0l)0}
   + \tilde{g} \left[ (1-K) \frac{\partial h_{00}}{\partial x^l}
   + K \frac{\partial h_{nn}}{\partial x^l} - \frac{\partial h_{ln}}{\partial x_n} \right] .
\end{equation}
All the above evolution equations are gauge invariant. These evolution equations suggest that also $K=0$ could be an appealing choice.

We now turn from the evolution equations to the constraints implied by the composition rule. The obvious constraints are obtained by choosing only spatial indices in Eq.~(\ref{glinA}),
\begin{equation}\label{glinconstraints1a}
   A^{(kl)}_j = \frac{1}{2} \left(
   \frac{\partial h_{jl}}{\partial x^k} - \frac{\partial h_{jk}}{\partial x^l} \right)
   + \frac{1}{2\tilde{g}} \, \frac{\partial \omega_{kl}}{\partial x^j} .
\end{equation}
Further constraints are obtained by considering $A_{(0k)l}-A_{(0l)k}$,
\begin{equation}\label{glinconstraints1b}
    A^{(0k)}_l - A^{(0l)}_k = \frac{1}{2} \left( \frac{\partial h_{0l}}{\partial x^k} -
   \frac{\partial h_{0k}}{\partial x^l} \right)
   + \frac{1}{2\tilde{g}} \left( \frac{\partial \omega_{0l}}{\partial x^k} -
   \frac{\partial \omega_{0k}}{\partial x^l} \right) .
\end{equation}

In total, we have turned the composition rule for the $24$ components of the gauge vector fields and the $4$ coordinate conditions (\ref{harmoniclin}) into the $16$ evolution equations (\ref{hklevol}), (\ref{omklevol}) and (\ref{h0kevol})--(\ref{om0kevol}) for the tetrad variables and the $9+3=12$ constraints (\ref{glinconstraints1a}), (\ref{glinconstraints1b}). We refer to these constraints resulting directly from the composition rule as the primary constraints of the composite theory. These primary constraints are not affected by coupling the gravitational field to matter. However, the evolution equations for the tetrad variables should be expected to be changed by coupling terms in the Hamiltonian.

\section{Pure field theory}\label{secpurefields}
We are now ready to define the canonical Hamiltonian version of the composite theory of gravity in the weak-field approximation on the combined space of Yang-Mills and tetrad fields, following the general ideas developed in \cite{hco237}. We first provide the Hamiltonian and then elaborate a number of its implications.

\subsection{Hamiltonian}
The Hamiltonian for the composite theory of pure gravity,
\begin{equation}\label{Hampurefields}
   H = H_{\rm YM} + H_{\rm YM/t} ,
\end{equation}
consists of two contributions describing the workhorse theory and reproducing the evolution equations obtained from the composition rule, respectively. Our workhorse theory is the linearized version of the Yang-Mills theory based on the Lorentz group on the space $(A_{a \mu}, E^{a \mu})$. The proper Hamiltonian is given by (see, e.g., Section~15.2 of \cite{PeskinSchroeder}, Chap.~15 of \cite{WeinbergQFT2}, or \cite{hco229}; a derivation from the Yang-Mills Lagrangian is given in Appendix~\ref{appL2H}),
\begin{eqnarray}
   H_{\rm YM} &=& \int \bigg[ \frac{1}{2} \left( E^{a \mu} E_{a \mu}
   + \frac{\partial A_{a i}}{\partial x^j} \frac{\partial A^{a i}}{\partial x_j}
   - \frac{\partial A_{a i}}{\partial x^j} \frac{\partial A^{a j}}{\partial x_i}
   \right) 
   	\nonumber \\
   &-& E^{a 0} \frac{\partial A_{a j}}{\partial x_j}
   - E^{a j} \frac{\partial A_{a 0}}{\partial x^j} \bigg]  d^3 x .
\label{pH2LHamf}
\end{eqnarray}
The Hamiltonian for coupling the Yang-Mills and tetrad variables,
\begin{eqnarray}
   H_{\rm YM/t} &=& \int \dot{b}_{\kappa\lambda} \, p^{\kappa\lambda} \, d^3 x
   \nonumber\\ &=& \frac{1}{4} \int \left(
   \frac{\partial h_{\kappa\lambda}}{\partial t} \, \tilde{h}^{\kappa\lambda}
   + \frac{\partial \omega_{\kappa\lambda}}{\partial t} \, \tilde{\omega}^{\kappa\lambda}
   \right) \, d^3 x , \qquad 
\label{pglinHcoupl}
\end{eqnarray}
is chosen such that the canonical evolution equations
\begin{equation}\label{Hamtetradeqs}
   \frac{\partial b_{\kappa\lambda}}{\partial t} = \frac{\delta H}{\delta p^{\kappa\lambda}} ,
   \qquad
   \frac{\partial p^{\kappa\lambda}}{\partial t} = - \frac{\delta H}{\delta b_{\kappa\lambda}} ,
\end{equation}
reproduce the evolution equations (\ref{hklevol}), (\ref{omklevol}) and (\ref{h0kevol})--(\ref{om0kevol}) for the tetrad variables. These evolution equations implied by the composition rule and the coordinate conditions (\ref{harmoniclin}) are of crucial importance for finding the Hamiltonian $H_{\rm YM/t}$, that is, for obtaining the complete canonical Hamiltonian formulation of the composite theory.

We have introduced the variables $p^{\kappa\lambda}$ in a purely formal manner as the conjugate momenta of the tetrad variables. At this point, we can offer a physical interpretation. Note that, in view of the evolution equations of the tetrad variables, the Hamiltonian $H_{\rm YM/t}$ contains a contribution that is bilinear in the variables $p^{\kappa\lambda}$ and the gauge vector fields $A_{a\mu}$. This contribution can be written in the form $-J^{a\mu} A_{a\mu}$ with the identifications
\begin{equation}\label{conjtetradinterpret0l}
   -J^{(0l)0} = \tilde{g} \tilde{\omega}^{0l} , \quad
   -J^{(0l)j} = \frac{1}{2} \tilde{h}^{lj} - \frac{1}{2} \frac{K}{1-K} \tilde{h}^{00} \eta^{lj} ,
\end{equation}
and
\begin{equation}\label{conjtetradinterpretkl}
   -J^{(kl)0} = \tilde{g} \tilde{\omega}^{kl} , \quad -J^{(kl)j} = 0 .
\end{equation}
The symmetric and antisymmetric parts of the variables $p^{\kappa\lambda}$ play the role of external Yang-Mills fluxes. By requiring Lorentz covariant fluxes, Eq.~(\ref{conjtetradinterpretkl}) immediately leads to the conclusion
\begin{equation}\label{tetradomtilkl}
   \tilde{\omega}^{kl} = 0 .
\end{equation}
Only the external fluxes $J^{(0l)\mu}$ can be nonvanishing in our composite Yang-Mills theory of gravity.

\subsection{Field equations}\label{secpurefieldeqs}
For the evolution of the conjugate momenta of the tetrad variables, we find the following results by means of Eq.~(\ref{Hamtetradeqs}):
\begin{eqnarray}
   \frac{\partial \tilde{h}^{kl}}{\partial t} &=&
   \frac{\partial (\tilde{h}^{0k} - \tilde{g} \tilde{\omega}^{0k})}{\partial x_l}
   + \frac{\partial (\tilde{h}^{0l} - \tilde{g} \tilde{\omega}^{0l})}{\partial x_k}
   \nonumber \\ &-& 2 K \delta_{kl} \,
   \frac{\partial (\tilde{h}^{0n} - \tilde{g} \tilde{\omega}^{0n})}{\partial x^n} ,
\label{htilevolkl}
\end{eqnarray}
\begin{equation}\label{htilevol00}
   \frac{\partial \tilde{h}^{00}}{\partial t} =
   2 K \, \frac{\partial \tilde{h}^{0l}}{\partial x^l}
   + 2 \tilde{g} \, (1-K) \, \frac{\partial \tilde{\omega}^{0l}}{\partial x^l} ,
\end{equation}
\begin{equation}\label{htilevol0l}
   \frac{\partial \tilde{h}^{0l}}{\partial t} = 
   \frac{1}{2} \frac{\partial \tilde{h}^{ln}}{\partial x^n}
   + \frac{1}{2} \frac{\partial \tilde{h}^{00}}{\partial x_l} ,
\end{equation}
\begin{equation}\label{omtilevol0l}
   \frac{\partial \tilde{\omega}^{0l}}{\partial t} = 
   - \frac{1}{2 \tilde{g}} \frac{\partial \tilde{h}^{ln}}{\partial x^n}
   + \frac{1}{2 \tilde{g}} \frac{K}{1-K} \, \frac{\partial \tilde{h}^{00}}{\partial x_l} ,
\end{equation}
and
\begin{equation}\label{omtilevolkl}
   \frac{\partial \tilde{\omega}^{kl}}{\partial t} = 0 .
\end{equation}
Note that these equations for the conjugate momenta of the tetrad variables are independent of any other variables. The last of these evolution equations is consistent with our previous conclusion (\ref{tetradomtilkl}). According to the definition (\ref{conjtetradinterpret0l}), Eq.~(\ref{omtilevol0l}) can be rewritten as
\begin{equation}\label{Jconservation}
   \frac{\partial J^{(0l)\mu}}{\partial x^\mu} = 0 ,
\end{equation}
which supports our interpretation of conjugate tetrad variables in terms of conserved fluxes.

The evolution equations for the Yang-Mills fields are obtained from (note the sign conventions)
\begin{equation}\label{HamYMeqs}
   \frac{\partial A_{a \mu}}{\partial t} = - \frac{\delta H}{\delta E^{a \mu}} ,
   \qquad
   \frac{\partial E^{a \mu}}{\partial t} = \frac{\delta H}{\delta A_{a \mu}} .
\end{equation}
The resulting equations can be written in the following form:
\begin{equation}\label{Aevola0}
   \frac{\partial A^a_0}{\partial t} = - E^a_0 + \frac{\partial A^a_n}{\partial x_n} ,
\end{equation}
and
\begin{equation}\label{Aevolaj}
   \frac{\partial A^a_j}{\partial t} = - E^a_j + \frac{\partial A^a_0}{\partial x^j} ,
\end{equation}
for the gauge vector fields, whereas their conjugate partners are governed by
\begin{equation}\label{Eevol0k0}
   \frac{\partial E^{(0l)}_0}{\partial t} = - \frac{\partial E^{(0l)}_n}{\partial x_n}
   - J^{(0l)}_0 ,
\end{equation}
\begin{equation}\label{Eevolkl0}
   \frac{\partial E^{(kl)}_0}{\partial t} = - \frac{\partial E^{(kl)}_n}{\partial x_n} ,
\end{equation}
\begin{equation}\label{Eevol0kj}
   \frac{\partial E^{(0l)}_j}{\partial t} = - \frac{\partial E^{(0l)}_0}{\partial x^j}
   - \frac{\partial^2 A^{(0l)}_j}{\partial x^n \partial x_n}
   + \frac{\partial^2 A^{(0l)}_n}{\partial x^j \partial x_n} - J^{(0l)}_j ,
\end{equation}
and
\begin{equation}\label{Eevolklj}
   \frac{\partial E^{(kl)}_j}{\partial t} = -\frac{\partial E^{(kl)}_0}{\partial x^j}
   - \frac{\partial^2 A^{(kl)}_j}{\partial x^n \partial x_n}
   + \frac{\partial^2 A^{(kl)}_n}{\partial x^j \partial x_n} .
\end{equation}
Note that these evolution equations are gauge invariant, provided that Eq.~(\ref{Lamevoleq}) for $\Lambda_a$ holds. These are the linearized standard field equations for Yang-Mills fields, which are strongly reminiscent of Maxwell's equations of electrodynamics.

Equation (\ref{Aevolaj}), together with the representation (\ref{glinA}), implies the useful identity
\begin{equation}\label{EklnJacobi}
    E^{(ln)}_k +  E^{(nk)}_l +  E^{(kl)}_n = 0 .
\end{equation}
This identity remains valid when we later include matter [that is, it can more generally be derived from Eq.~(\ref{representEklj})].

\subsection{Constraints}
The primary constraints (\ref{glinconstraints1a}), (\ref{glinconstraints1b}) must be valid at all times. From the time derivative of the primary constraints we obtain secondary constraints, a further time derivative yields tertiary constraints, and so on. This iterative process, in which the required time derivatives are evaluated by means of the evolution equations, is continued until no further constraints arise. The crucial question is whether the iterative process stops before all degrees of freedom are fixed by constraints. As the introduction revealed that, in the canonical Hamiltonian formulation of composite gravity, we are dealing with $80$ fields, we need around $75$ constraints to obtain an appropriate number of degrees of freedom for a theory of gravity.

The secondary constraints obtained as the time derivatives of the primary constraints can be formulated nicely in terms of Yang-Mills variables,
\begin{equation}\label{glinconstraints2a}
   E^{(kl)}_j = \frac{\partial A^{(0j)}_l}{\partial x^k} - \frac{\partial A^{(0j)}_k}{\partial x^l} , 
\end{equation}
\begin{equation}\label{glinconstraints2b}
   E^{(0l)}_k = E^{(0k)}_l ,
\end{equation}
and the tertiary constraints are subsequently obtained as
\begin{equation}\label{glinconstraints3a}
   \frac{\partial E^{(kl)}_0}{\partial x^j} - \frac{\partial E^{(0j)}_l}{\partial x^k}
   + \frac{\partial E^{(0j)}_k}{\partial x^l} = \frac{\partial}{\partial x_n}
   \left( \frac{\partial A^{(kl)}_n}{\partial x^j} - \frac{\partial A^{(kl)}_j}{\partial x^n} \right) ,
\end{equation}
\begin{equation}\label{glinconstraints3b}
   \frac{\partial E^{(0l)}_0}{\partial x_k} - \frac{\partial E^{(0k)}_0}{\partial x_l} =
   \frac{\partial E^{(kl)}_n}{\partial x_n} .
\end{equation}
The latter constraint has been simplified by means of the identity (\ref{EklnJacobi}). Note that these tertiary constraints can be used to rewrite the evolution equations (\ref{Eevolkl0}) and (\ref{Eevolklj}) as
\begin{equation}\label{Eevolkl0cx}
   \frac{\partial E^{(kl)}_0}{\partial t} = \frac{\partial E^{(0k)}_0}{\partial x^l}
   - \frac{\partial E^{(0l)}_0}{\partial x^k} ,
\end{equation}
and
\begin{equation}\label{Eevolkljcx}
   \frac{\partial E^{(kl)}_j}{\partial t} =
   \frac{\partial E^{(0j)}_k}{\partial x^l} - \frac{\partial E^{(0j)}_l}{\partial x^k} .
\end{equation}

Up to this point, the variables $p^{\kappa\lambda}$ do not appear in the constraints. From now on, only the variables $p^{\kappa\lambda}$ occur in the constraints. In the next round, we find
\begin{equation}\label{glinconstraints4a}
   \frac{\partial J^{(0l)j}}{\partial x_k}
   - \frac{\partial J^{(0k)j}}{\partial x_l} = 0 ,
\end{equation}
\begin{equation}\label{glinconstraints4b}
   \frac{\partial J^{(0l)0}}{\partial x_k}
   - \frac{\partial J^{(0k)0}}{\partial x_l} = 0 .
\end{equation}
As we assume that, in the absence of matter, the external fluxes (\ref{conjtetradinterpret0l}) vanish, these last conditions are satisfied trivially so that the hierarchy of constraints ends at this point.

We have arrived at a total of $4 \times 12 = 48$ constraints resulting from the composition rule, supplemented by the three constraints (\ref{tetradomtilkl}) so that the total is $51$. All these constraints are gauge invariant. This is a consequence of the fact that the composition rule is designed such that the four-vector fields $A_{a \nu}$ possess the proper gauge transformation behavior (\ref{gaugeA}) and all the evolution equations are gauge invariant. We eventually argue in favor of the $16$ constraints $p^{\kappa\lambda}=0$ (or $\tilde{h}^{\kappa\lambda} = \tilde{\omega}^{\kappa\lambda} = 0$), which would replace the $15$ constraints (\ref{tetradomtilkl}), (\ref{glinconstraints4a}), (\ref{glinconstraints4b}) and actually increase the count by one.

In a Yang-Mills theory, half of the degrees of freedom can be eliminated by gauge constraints (roughly speaking, the four-vector potentials have only transverse components, no longitudinal or temporal ones). In our case, we have $24$ gauge constraints, which brings us to a total of $75$ (or $76$) constraints for our $80$ fields. It is quite remarkable that just a few of the $80$ degrees of freedom survive, as we would expect for a theory of gravity.

\subsection{Compact form of theory}\label{seccompactformpure}
The goal of this subsection is to find a closed set of differential equations for the tetrad variables. To reach this goal it is important to express all the Yang-Mills variables in terms of the tetrad variables. For the vector fields $A_{a \mu}$, the desired expression is given by the composition rule (\ref{glinA}). Their conjugates $E^{a \mu}$ can then be extracted from the evolution equations (\ref{Aevola0}), (\ref{Aevolaj}) (see Appendix \ref{appYMtetrad} for a summary of the resulting expressions).

As we have already recognized $J^{(0l)\mu} = 0 = \tilde{\omega}^{kl}$, Eqs.~(\ref{htilevolkl})--(\ref{omtilevolkl}) imply that all conjugate tetrad variables must be constant and can be assumed to be zero,
\begin{equation}\label{tetradallmom}
   p^{\kappa\lambda} = 0 .
\end{equation}
This is a very desirable condition for the natural canonical Hamiltonian approach to composite theories. As the conjugate momenta $p^{\kappa\lambda}$ appear linearly in the Hamiltonian (\ref{pglinHcoupl}), they lead to an unbounded Hamiltonian and consequently to the famous risk of instabilities in higher derivative theories \cite{Ostrogradsky1850,Woodard15}. Avoiding such instabilities is an important topic, in particular, in alternative theories of gravity \cite{Chenetal13,RaidalVeermae17,Stelle77,Stelle78,Krasnikov87,GrosseKnetter94,Beckeretal17,Salvio19}. The constraints (\ref{tetradallmom}) provide the most obvious way of eliminating instabilities in the canonical Hamiltonian approach to composite higher derivative theories, which differs from the usual Ostrogradsky approach \cite{hco235,hco237}. Moreover, this constraint implies that we have to solve the original Yang-Mills equations without any modification. In other words, the composite theory simply selects solutions of the Yang-Mills theory based on the Lorentz group to obtain the composite theory of gravity. This insight provides a more direct argument for the stability of solutions. Note that the large number of constraints and the small number of remaining degrees of freedom indicates that the composite theory is highly selective.

From Eq.~(\ref{Eevolklj}) we obtain
\begin{equation}\label{tetradhkl}
   \frac{\partial^2 h_{kl}}{\partial x_\mu \partial x^\mu} =
   \frac{\partial^2 f}{\partial x^k \partial x^l} ,
\end{equation}
where the unknown function $f$ results from integration, and similarly Eq.~(\ref{Eevolkl0}) gives
\begin{equation}\label{tetradh0l}
   \frac{\partial^2 h_{0l}}{\partial x_n \partial x^n} = \frac{\partial f_0}{\partial x^l} .
\end{equation}
Equations (\ref{Eevol0k0}) and (\ref{Eevol0kj}) provide further integrability conditions that can be exploited in a similar manner. Consolidating all the results, we get the following compact formula summarizing the linear version of the composite theory of pure gravity on the level of tetrad variables, 
\begin{equation}\label{tetradhall}
   \frac{\partial^2 h_{\mu\nu}}{\partial x_\lambda \partial x^\lambda} =
   \frac{\partial^2 f}{\partial x^\mu \partial x^\nu} ,
\end{equation}
possibly after a minor redefinition of $f$.

An interesting feature of these field equations is that the function $f$, which results from integration, needs to be determined simultaneously with the solutions $h_{\mu\nu}$. We arrive at a set of second-order differential equations because higher derivative equations play the role of integrability conditions. The coupling constant $\tilde{g}$ does not occur in these equations. Possible antisymmetric contributions to the tetrad variables are governed by the wave equations (\ref{tetradom}), and all conjugate momenta of the tetrad variables must vanish according to Eq.~(\ref{tetradallmom}).

\subsection{Comparison to general relativity}
Einstein's field equation for pure gravity in the weak-field approximation to general relativity is given by a vanishing curvature tensor [see Eq.~(\ref{linRt})],
\begin{equation}\label{GRgeneq}
   \frac{\partial^2 h_{\mu\nu}}{\partial x_\lambda \partial x^\lambda}
   - \frac{\partial^2 {h^\lambda}_\mu}{\partial x^\lambda \partial x^\nu}
   - \frac{\partial^2 {h^\lambda}_\nu}{\partial x^\mu \partial x^\lambda}
   + \frac{\partial^2 {h^\lambda}_\lambda}{\partial x^\mu \partial x^\nu} = 0 .
\end{equation}
It is important to note that the coordinates $x^\mu$ in general relativity are not associated with an underlying Minkowski space so that these field equations can be simplified by suitable general coordinate transformations. If we impose the same coordinate conditions (\ref{harmoniclin}) as used in composite gravity, the field equations (\ref{GRgeneq}) of linearized general relativity simplify to
\begin{equation}\label{GRgeneq1}
   \frac{\partial^2 h_{\mu\nu}}{\partial x_\lambda \partial x^\lambda} =
   (2K-1) \frac{\partial^2 {h^\lambda}_\lambda}{\partial x^\mu \partial x^\nu} .
\end{equation}
This equation coincides with Eq.~(\ref{tetradhall}) for composite gravity for $f=(2K-1) {h^\lambda}_\lambda$. It becomes particularly simple for harmonic coordinates with $K=1/2$, which may be pictured as nearly Minkowskian (see, e.g., pp.\,163 and 254 of \cite{Weinberg}).

As in general relativity, the solutions for the deviatoric metric $h_{\mu\nu}$ in harmonic coordinates can assume all kinds of polarization states, including longitudinal and temporal components. The actual polarization of gravitational waves depends on the nature of their source (typically binary systems of two black holes, two neutron stars, or a black hole and a neutron star during their in-spiral or merger phases).

\subsection{Static isotropic solution}\label{secisostatsol}
To find the static isotropic solution for the weak-field approximation to composite gravity for the coordinate conditions (\ref{harmoniclin}), we start from the general ansatz
\begin{equation}\label{linisostath}
   h_{00} = \bar{\beta}(r) , \quad
   h_{kl} = \bar{\alpha}(r) \delta_{kl}  + \bar{\xi}(r) \frac{x_k x_l}{r^2} , \quad
   h_{0k} = h_{k0} = 0 ,
\end{equation}
with $r=(x_1^2+x_2^2+x_3^2)^{1/2}$. The coordinate conditions (\ref{harmoniclin}) become
\begin{equation}\label{harmonicaltisolin}
   r \Big[ (3K-1) \bar{\alpha}'- K \bar{\beta}' + (K-1) \bar{\xi}' \Big] = 2 \bar{\xi} .
\end{equation}
A prime on a function of $r$ indicates the derivative with respect to $r$.

We assume that also $f$ in Eq.~(\ref{tetradhall}) is static and isotropic.
With $f=f(r)$, Eq.~(\ref{tetradhall}) leads to two equations,
\begin{equation}\label{linbetadiffeq}
   r^2 \bar{\beta}'' + 2 r \bar{\beta}' = 0 ,
\end{equation}
and
\begin{eqnarray}
   \frac{x_k x_l}{r^2} \left( r^2 \bar{\xi}'' + 2r \bar{\xi}' - 6 \bar{\xi} -r^2 f'' +r f'
   \right) &=& \nonumber \\ && \hspace{-10em}
   \delta_{kl} \left( r f' - 2 \bar{\xi} -r^2 \bar{\alpha}'' -2 r \bar{\alpha}' \right) ,
   \qquad 
\label{sepcompaz}
\end{eqnarray}
where each side of the latter equation must vanish separately.

All these equations are of the equidimensional type, that is, in each term there are as many factors of $r$ as there are derivatives with respect to $r$, suggesting simple power-law solutions. Equation~(\ref{linbetadiffeq}) implies $\bar{\beta}' \propto r^{-2}$ and we hence write
\begin{equation}\label{linbetasol}
   \bar{\beta}(r) = 2 \, \frac{r_0}{r} ,
\end{equation}
where $r_0$ is a constant length scale and a possible additive constant has been omitted to obtain asymptotic Minkowskian behavior. Equation (\ref{harmonicaltisolin}) suggests that $\bar{\alpha}$ has the same power-law decay, so that the right-hand side of Eq.~(\ref{sepcompaz}) implies $r f' = 2 \bar{\xi}$. Equation (\ref{harmonicaltisolin}) provides a relation among prefactors, so that we can write
\begin{equation}\label{linalphaxisol}
   \bar{\alpha}(r) = \frac{\bar{c}}{1-K} \, \frac{r_0}{r} , \qquad
   \bar{\xi}(r) = \frac{\bar{c}-(3\bar{c}-2)K-2K^2}{1-K^2} \, \frac{r_0}{r} .
\end{equation}
Consistency with general relativity, which implies a vanishing curvature tensor, requires $\bar{c}=1$. The condition $\bar{c}=1$ is not predicted by the weak-field approximation of pure composite gravity, but it arises naturally in the full, nonlinear theory  or from a suitable coupling to matter (see Section~\ref{seccompactmat} below).

\section{Coupling of field to matter}\label{secwithmatter}
Of course, we cannot really appreciate a theory of the gravitational field without coupling it to matter. On the one hand, we want to understand the gravitational field generated by matter, say for calculating the parameters $\bar{c}$ and $r_0$ in the solution given in Eq.~(\ref{linalphaxisol}). On the other hand, we want to understand the motion of matter in a gravitational field.

The most convenient options for describing matter are given by point particle mechanics or hydrodynamics. We here consider a single point particle either generating a gravitational field or moving in a gravitational field.

\subsection{Particle in a gravitational field}
As a starting point for discussing particle motion in a weak gravitational field, we use the first-order expansion of the standard Hamiltonian,
\begin{equation}\label{weakfieldparticleH}
   H_{\rm m} = \gamma m - \frac{1}{2 \gamma m} \, p_\mu p_\nu h^{\mu\nu} ,
\end{equation}
where $m$ is the rest mass of the particle, $h^{\mu\nu}$ depends on the particle position $x^j$, the particle momentum is given by $p_j$, and we define $-p_0=p^0=m \gamma$, where
\begin{equation}\label{sprelgammap}
   \gamma = \left[ 1 + \left( \frac{\bm{p}}{m} \right)^2 \right]^{\frac{1}{2}} ,
\end{equation}
is a function of the spatial components $p_j$ of the particle momentum. When a higher order definition of $p^0$ is required, one should use $p^0=H_{\rm m}$, where Eq.~(\ref{weakfieldparticleH}) provides the first-order result in $h^{\mu\nu}$. The lowest-order energy-momentum tensor is given by (see, e.g., Eq.~(2.8.4) of \cite{Weinberg})
\begin{eqnarray}
   T_{\mu\nu} &=& -2 \frac{\delta H_{\rm m}}{\delta h^{\mu\nu}} 
   = \frac{p_\mu p_\nu}{\gamma m} \, \delta^3(\bm{x}-\bm{x}(t))
   \nonumber \\
   &=& \gamma m \frac{d x_\mu}{d t} \frac{d x_\nu}{d t} \, \delta^3(\bm{x}-\bm{x}(t)) ,
\label{weakfieldparticleT}
\end{eqnarray}
where the lowest-order result $p_j=m \gamma \, dx_j/dt$ has been used. The evolution equation $dp_j/dt=0$ for a free particle in the absence of gravity leads to the result
\begin{equation}\label{energymomtimeder}
   \frac{\partial T_{\mu\nu}}{\partial t} =
   - \frac{\partial T_{\mu\nu}}{\partial x^j} \, \frac{d x^j}{d t} =
   \frac{p_j}{p_0} \, \frac{\partial T_{\mu\nu}}{\partial x^j} ,
\end{equation}
from which, for $\nu=0$, we obtain energy-momentum conservation in the form
\begin{equation}\label{energymomconsmattT}
   \frac{\partial T^{\mu\nu}}{\partial x^\nu} = 0 .
\end{equation}

By construction, the Hamiltonian $(\ref{weakfieldparticleH})$ leads to geodesic motion in a weak field. The potential distortion of geodesic motion by further couplings between matter and field is explored in Section \ref{secmodpartmot} below.

\subsection{Hamiltonian for coupling field and matter}
The occurrence of $h^{\mu\nu}$ in the Hamiltonian (\ref{weakfieldparticleH}) already implies a coupling of field and matter. It leads to geodesic motion in the given field $h^{\mu\nu}$, but it does not provide meaningful field equations for determining gravitational fields. For that purpose we need to couple the Yang-Mills field to the energy-momentum tensor of matter. In Appendix~\ref{appL2H}, the details of the coupling are discussed in a Lagrangian setting, and the following Hamiltonian for the coupling is obtained,
\begin{equation}\label{Hcoupling}
   H_{\rm YM/m} = \int \left( F^{(\lambda n)}_{jn} {C^j}_\lambda - E^{(\lambda j)}_j {C^0}_\lambda
   - E^{(0 l)}_j {C^j}_l \right) d^3x ,
\end{equation}
with
\begin{equation}\label{Ctensorchoice0}
   C_{\mu\nu} = G_1 \, \mathring{T}_{\mu\nu} + G_2 \, \eta_{\mu\nu}  {T^\lambda}_\lambda ,
\end{equation}
where $\mathring{T}_{\mu\nu}$ is the traceless part of the energy-momentum tensor of matter defined in Eq.~(\ref{Tringdef}) and the coefficients $G_1$, $G_2$ must have the same dimensions as Newton's constant $G$ (cf.\ Table~\ref{tabledimensions}). The concrete values of  $G_1$, $G_2$ can only be chosen once we have elaborated all the equations for gravitational fields coupled to matter.

\begin{table}
\begin{tabular}{cc}
   %\hline
   Quantities & Dimensions \\
   \hline
   $g_{\mu\nu}$, ${b^\kappa}_\mu$, $h_{\mu\nu}$, $\omega_{\mu\nu}$, $\Lambda_{\rm E}$ & --- \\
   $A^a_\nu$, $H$ & $L^{-1}$ \\
   $E^a_\nu$, $B^a_\nu$, $F^a_{\mu\nu}$, $R^{\mu\nu}$  & $L^{-2}$ \\
   $p^{\kappa\lambda}$, $\tilde{h}^{\kappa\lambda}$, $\tilde{\omega}^{\kappa\lambda}$  & $L^{-3}$ \\
   $V_{\kappa\lambda}$ & $M$ \\
   $T_{\mu\nu}$ & $L^{-3} M$ \\
   $G$, $G_1$, $G_2$  & $L M^{-1}$ \\
   %\hline
\end{tabular}
\caption{Dimensions of various quantities in terms of length ($L$) and mass ($M$) for $c=1$.}
\label{tabledimensions}
\end{table}

The Lagrangian associated with the Hamiltonian (\ref{Hcoupling}) for the coupling of the Yang-Mills field to matter has previously been proposed in Eqs.~(51) and (52) of \cite{hco231}. We here introduce an additional coupling of the tetrad field to matter,
\begin{equation}\label{Htetradmatter}
   H_{\rm t/m} = \int V_{\kappa\lambda} \, p^{\kappa\lambda} \, d^3x ,
\end{equation}
where, for the linearized theory, $V_{\kappa\lambda}$ can be assumed to be a symmetric tensor to be constructed from the energy-momentum tensor of matter (more precisely, the time derivative of $V_{\kappa\lambda}$ turns out to be a tensor; the defining equations and more insight into the role of the indices are provided in Section~\ref{secmodifiedconstr}). The idea behind this additional coupling is as follows. In the absence of matter, the composite theory selects solutions from a pure Yang-Mills theory, which is a consequence of the vanishing conjugate momenta $p^{\kappa\lambda}$ of the tetrad variables established in Eq.~(\ref{tetradallmom}). In the presence of matter, it is more natural to select solutions of the Yang-Mills theory with suitable external fluxes, so that the conjugate momenta $p^{\kappa\lambda}$ should no longer be expected to vanish. Use of the separate Hamiltonian (\ref{Htetradmatter}) in addition to the previously suggested coupling mechanism (\ref{Hcoupling}) allows us to find a consistently tuned coupling of both Yang-Mills and tetrad fields to matter. Note that the ``general wisdom'' about the possibilities of coupling gravity to matter \cite{Feynmanetal,Deser70,Straumann00} is not beyond all doubt \cite{Padmanabhan08} and, in the context of composite theories, this coupling can be even richer.

For obtaining the composite theory of gravity in the presence of matter, we would like to add the Hamiltonians $H_{\rm YM/m}$, $H_{\rm t/m}$ and $H_{\rm m}$ introducing the coupling of field and matter to the Hamiltonian (\ref{Hampurefields}) of pure gravity. However, there is a problem. With the help of Table~\ref{tabledimensions}, we realize that the Hamiltonian (\ref{Hampurefields}) has dimension of ${\rm length}^{-1}$, and so does the Hamiltonian $H_{\rm YM/m}$ defined in Eq.~(\ref{Hcoupling}). The dimensions of the Hamiltonian $H_{\rm t/m}$ defined in Eq.~(\ref{Htetradmatter}) can still be adjusted by the definition of $V_{\kappa\lambda}$. However, the Hamiltonian $H_{\rm m}$ defined in Eq.~(\ref{weakfieldparticleH}) has dimensions of mass, which is what we actually expect for a Hamiltonian when using the speed of light as the unit for velocities ($c=1$).

As the mismatch in dimensions can be regarded as an action factor, it seems natural to multiply $H_{\rm YM} + H_{\rm YM/t} + H_{\rm YM/m}$ by Planck's constant $\hbar$. We do that implicitly by using $\hbar$ as the unit of action ($\hbar=1$), thus eliminating the dimensional mismatch. However, this choice of units implies that, in $H_{\rm YM} + H_{\rm YM/t}$, we actually deal with the energy of gravitational field quanta, which is clearly not the most appropriate energy scale when we usually consider problems involving gravity. We hence introduce a very small dimensionless parameter $\Lambda_{\rm E}$ to scale down the typical energy associated with gravitationally interacting masses to the level of graviton energies,
\begin{equation}\label{Hamfullsystem}
   H = H_{\rm YM} + H_{\rm YM/t} + H_{\rm YM/m} + \Lambda_{\rm E} (H_{\rm t/m} + H_{\rm m}) .
\end{equation}
In the Lagrangian formulation in Eq.~(52) of \cite{hco231}, it can be recognized that $\Lambda_{\rm E}$ plays the role of a dimensionless cosmological constant in general relativity. We hence write
\begin{equation}\label{LambdaEdef}
   \Lambda_{\rm E} = \left( \frac{\ell_{\rm p}}{D} \right)^2 ,
\end{equation}
where $\ell_{\rm p} = \sqrt{\hbar G/c^3} = \sqrt{G}$ is the Planck length and $D$ is the diameter of the observable universe. This parameter $\Lambda_{\rm E}$ can be estimated to be of the order of $10^{-124}$.
% D = 8.8×10^26 m --- https://en.wikipedia.org/wiki/Observable_universe#Size
% l_p = 1.616255(18)×10^−35 m --- https://en.wikipedia.org/wiki/Planck_length
It is interesting to note that even our formulation of classical gravity requires an action constant. A similar situation arises in formulating the entropy of a classical ideal gas, indicating that a deeper understanding of an ideal gas requires quantum theory. The same conclusion may be true for a deeper understanding of gravity.

\subsection{Modified field equations}
In the presence of matter, the dynamic aspects of the composition rule (\ref{glinA}) are affected by the Hamiltonian $H_{\rm t/m}$, but not its static aspects. In other words, the primary constraints are unchanged whereas the evolution equations for the tetrad variables get modified. As $V_{\kappa\lambda}$ is assumed to be symmetric, only the evolution equations (\ref{hklevol}), (\ref{h0kevol}) and (\ref{h00evol}) for $h_{\kappa\lambda}$ get changed,
\begin{eqnarray}
   \frac{\partial h_{kl}}{\partial t} &=& \frac{1}{2} \left( \frac{\partial h_{0l}}{\partial x^k}
   + \frac{\partial h_{0k}}{\partial x^l} \right) + A_{(0k)l} + A_{(0l)k} \nonumber \\
   &-& \frac{1}{2\tilde{g}} \left( \frac{\partial \omega_{0l}}{\partial x^k}
   + \frac{\partial \omega_{0k}}{\partial x^l} \right) + 2 \Lambda_{\rm E} V_{kl} , \qquad 
\label{hklevolmat}
\end{eqnarray}
\begin{equation}\label{h0kevolmat}
   \frac{\partial  h_{0l}}{\partial t} = \frac{\partial h_{ln}}{\partial x_n}
   - K \, \frac{\partial {h^\nu}_\nu}{\partial x^l} + 2 \Lambda_{\rm E} V_{0l} ,
\end{equation}
and
\begin{equation}\label{h00evolmat}
   \frac{\partial  h_{00}}{\partial t} = \frac{\partial h_{0l}}{\partial x_l} 
   - \frac{K}{1-K} \left[ 2 A_{(0l)l}
   - \frac{1}{\tilde{g}} \frac{\partial \omega_{0l}}{\partial x_l} \right]
   + 2 \Lambda_{\rm E} V_{00} .
\end{equation}
Equations~(\ref{h0kevolmat}) and (\ref{h00evolmat}) imply a tiny modification of the coordinate conditions (\ref{harmoniclin}).

The Hamiltonian $H_{\rm YM/m}$ given in Eq.~(\ref{Hcoupling}) depends only on the spatial components of the Yang-Mills fields, so that the evolution equations (\ref{Aevola0}), (\ref{Eevol0k0}) and (\ref{Eevolkl0}) for the temporal components of the Yang-Mills fields remain unaffected. Equation (\ref{Aevolaj}) gets modified to
\begin{equation}\label{Aevol0kjmod}
   \frac{\partial A^{(0l)}_j}{\partial t} = - E^{(0l)}_j + \frac{\partial A^{(0l)}_0}{\partial x^j}
   - C_{jl} + \delta_{jl} \, C_{00} ,
\end{equation}
and
\begin{equation}\label{Aevolkljmod}
   \frac{\partial A^{(kl)}_j}{\partial t} = - E^{(kl)}_j + \frac{\partial A^{(kl)}_0}{\partial x^j}
   + \delta_{jk} C_{0l} - \delta_{jl} C_{0k} .
\end{equation}
whereas Eqs.~(\ref{Eevol0kj}) and (\ref{Eevolklj}) become
\begin{eqnarray}
   \frac{\partial E^{(0l)}_j}{\partial t} &=& - \frac{\partial E^{(0l)}_0}{\partial x^j}
   - \frac{\partial^2 A^{(0l)}_j}{\partial x^n \partial x_n}
   + \frac{\partial^2 A^{(0l)}_n}{\partial x^j \partial x_n} - J^{(0l)}_j \qquad \nonumber \\
   &-& \frac{\partial C_{j0}}{\partial x_l} + \delta_{jl} \, \frac{\partial C_{n0}}{\partial x_n} , 
\label{Eevol0kjmod}
\end{eqnarray}
and
\begin{eqnarray}
   \frac{\partial E^{(kl)}_j}{\partial t} &=& -\frac{\partial E^{(kl)}_0}{\partial x^j}
   - \frac{\partial^2 A^{(kl)}_j}{\partial x^n \partial x_n}
   + \frac{\partial^2 A^{(kl)}_n}{\partial x^j \partial x_n} \nonumber \\
   &+& \frac{\partial C_{jk}}{\partial x^l} - \frac{\partial C_{jl}}{\partial x^k}
   + \delta_{jk} \, \frac{\partial C_{nl}}{\partial x_n}
   - \delta_{jl} \, \frac{\partial C_{nk}}{\partial x_n} . \qquad
\label{Eevolkljmod}
\end{eqnarray}
The fact that $C_{\mu\nu}$ occurs in Eqs.~(\ref{Aevol0kjmod}) and (\ref{Aevolkljmod}) for the gauge vector fields underlines that the coupling of the stress tensor to the workhorse theory of composite gravity does not happen via the usual flux mechanism for Yang-Mills theories.

The occurrence of $h^{\mu\nu}$ in Eq.~(\ref{weakfieldparticleH}) implies that the evolution equations (\ref{htilevolkl})--(\ref{htilevol0l}) for the symmetrized conjugate momenta $\tilde{h}^{\kappa\lambda}$ get modified, too. We find
\begin{eqnarray}
   \frac{\partial \tilde{h}^{kl}}{\partial t} &=&
   \frac{\partial (\tilde{h}^{0k} - \tilde{g} \tilde{\omega}^{0k})}{\partial x_l}
   + \frac{\partial (\tilde{h}^{0l} - \tilde{g} \tilde{\omega}^{0l})}{\partial x_k}
   \nonumber \\ &-& 2 K \delta_{kl} \,
   \frac{\partial (\tilde{h}^{0n} - \tilde{g} \tilde{\omega}^{0n})}{\partial x^n}
   + 2 \Lambda_{\rm E} \, T^{kl} ,
\label{htilevolklm}
\end{eqnarray}
\begin{equation}\label{htilevol00m}
   \frac{\partial \tilde{h}^{00}}{\partial t} =
   2 K \, \frac{\partial \tilde{h}^{0l}}{\partial x^l}
   + 2 \tilde{g} \, (1-K) \, \frac{\partial \tilde{\omega}^{0l}}{\partial x^l}
   + 2 \Lambda_{\rm E} \, T^{00} ,
\end{equation}
and
\begin{equation}\label{htilevol0km}
   \frac{\partial \tilde{h}^{0l}}{\partial t} = 
   \frac{1}{2} \frac{\partial \tilde{h}^{ln}}{\partial x^n}
   + \frac{1}{2} \frac{\partial \tilde{h}^{00}}{\partial x_l}
   + 2 \Lambda_{\rm E} \, T^{0l} .
\end{equation}
The occurrence of the energy-momentum tensor in Eqs.~(\ref{htilevolklm})--(\ref{htilevol00m}) is a very important qualitative modification. As anticipated, the conjugate momenta of the tetrad variables do not vanish in the presence of matter. Remember, however, that the dimensionless parameter $\Lambda_{\rm E}$ is extremely small.

\subsection{Modified constraints}\label{secmodifiedconstr}
In the presence of matter, the primary constraints (\ref{glinconstraints1a}) and (\ref{glinconstraints1b}) remain unchanged. The secondary constraints (\ref{glinconstraints2a}) change to
\begin{eqnarray}
   E^{(kl)}_j &=& \frac{\partial A^{(0j)}_l}{\partial x^k} - \frac{\partial A^{(0j)}_k}{\partial x^l}
   - \Lambda_{\rm E} \left( \frac{\partial V_{jl}}{\partial x^k}
   - \frac{\partial V_{jk}}{\partial x^l} \right) \nonumber \\
   &+& \delta_{jk} \, C_{0l} -  \delta_{jl} \, C_{0k} , \quad
\label{glinconstraints2amod}
\end{eqnarray}
whereas Eq.~(\ref{glinconstraints2b}) becomes
\begin{equation}\label{glinconstraints2bmod}
   E^{(0l)}_k - E^{(0k)}_l = \Lambda_{\rm E} \left( \frac{\partial V_{0l}}{\partial x^k}
   - \frac{\partial V_{0k}}{\partial x^l} \right) .
\end{equation}
The tertiary constraints (\ref{glinconstraints3a}) become
\begin{eqnarray}
   \frac{\partial E^{(kl)}_0}{\partial x^j} - \frac{\partial E^{(0j)}_l}{\partial x^k}
   + \frac{\partial E^{(0j)}_k}{\partial x^l} &=& \frac{\partial}{\partial x_n}
   \left( \frac{\partial A^{(kl)}_n}{\partial x^j} - \frac{\partial A^{(kl)}_j}{\partial x^n} \right)
   \nonumber \\
   && \hspace{-12em} + \Lambda_{\rm E} \frac{\partial}{\partial t}
   \left( \frac{\partial V_{jl}}{\partial x^k}
   - \frac{\partial V_{jk}}{\partial x^l} \right)
   + \delta_{jk} \, G_1 \frac{\partial T_{00} }{\partial x^l} 
   -  \delta_{jl} \, G_1 \frac{\partial T_{00} }{\partial x^k} ,
   \nonumber \\
\label{glinconstraints3amod}
\end{eqnarray}
and Eq.~(\ref{glinconstraints3b}) changes to
\begin{equation}\label{glinconstraints3bmod}
   \frac{\partial E^{(0l)}_0}{\partial x_k} - \frac{\partial E^{(0k)}_0}{\partial x_l} =
   \frac{\partial E^{(kl)}_n}{\partial x_n} 
   + \Lambda_{\rm E} \frac{\partial}{\partial x^\mu}
      \left( \frac{\partial V^{\mu l}}{\partial x_k}
      - \frac{\partial V^{\mu k}}{\partial x_l} \right) .
\end{equation}
Finally, the quaternary constraints (\ref{glinconstraints4a}) and (\ref{glinconstraints4b}) become
\begin{equation}\label{glinconstraints4amod}
   \frac{\partial J^{(0l)j}}{\partial x_k}
   - \frac{\partial J^{(0k)j}}{\partial x_l} =
   - \Lambda_{\rm E} \frac{\partial^2}{\partial x_\mu \partial x^\mu}
   \left( \frac{\partial V^{jl}}{\partial x_k}
   - \frac{\partial V^{jk}}{\partial x_l} \right) ,
\end{equation}
and
\begin{equation}\label{glinconstraints4bmod}
   \frac{\partial J^{(0l)0}}{\partial x_k}
   - \frac{\partial J^{(0k)0}}{\partial x_l} = 
   - \Lambda_{\rm E} \frac{\partial^2}{\partial x_\mu \partial x^\mu}
   \left( \frac{\partial V^{0l}}{\partial x_k}
   - \frac{\partial V^{0k}}{\partial x_l} \right) .
\end{equation}

At this stage we have to make a proper choice of the functions $V^{\kappa\lambda}$ in the Hamiltonian in order to avoid further constraints that would quickly make it impossible to find any solutions to the entire set of constraints. For this purpose we added a coupling of matter to the tetrad variables in addition to the more obvious coupling to the Yang-Mills variables. As a first step, we want to identify further vanishing conjugate tetrad variables because, according to Eq.~(\ref{conjtetradinterpret0l}), only the variables $\tilde{\omega}^{0l}$ and $\tilde{h}^{kl}$ carry essential information. Careful inspection of the structure of the evolution equations suggests the following choices of vanishing variables in addition to those given in Eq.~(\ref{tetradomtilkl}),
\begin{equation}\label{tetradconjsum0}
    \tilde{h}^{00} = 0 , \qquad  \tilde{h}^{0l} - \tilde{g} \tilde{\omega}^{0l} = 0 .
\end{equation}
The evolution equations for the conjugate tetrad variables then reduce to the much simpler form
\begin{equation}\label{htilevolklmX}
   \frac{\partial \tilde{h}^{kl}}{\partial t} = 2 \Lambda_{\rm E} \, T^{kl} , \qquad 
   \frac{\partial \tilde{h}^{kl}}{\partial x^l} = - 2 \Lambda_{\rm E} \, T^{k0} ,
\end{equation}
and
\begin{equation}\label{omtilevol0lX}
   \tilde{g} \, \frac{\partial \tilde{\omega}^{0l}}{\partial t} = 
   \Lambda_{\rm E} \, T^{0l} , \qquad 
   \tilde{g} \, \frac{\partial \tilde{\omega}^{0l}}{\partial x^l} =
   - \Lambda_{\rm E} \, T^{00} .
\end{equation}
Note that the consistency between the two members of each equation is guaranteed by energy-momentum conservation.

The quaternary constraints can now be satisfied if we construct $V^{\kappa\lambda}$ by solving the Poisson equations
\begin{equation}\label{tetradhomtilrepsm1}
   J^{(0l)\nu} = - \Lambda_{\rm E} \frac{\partial^2 V^{l\nu}}{\partial x_\mu \partial x^\mu} ,
\end{equation}
where suitable initial and boundary conditions need to be imposed to find $V^{l\nu}$. There is no need to choose any particular form of $V^{00}$ because, according to Eq.~(\ref{tetradhomtilrepsm1}), there is no flux component associated with it. We hence assume $V^{00}=0$, unless there is any particular need to modify Eq.~(\ref{h00evolmat}). Note that $\nu$ is a four-vector index whereas $l$ is related to the labels of the Lie algebra (more precisely, $l$ is the label for the Lorentz boosts). Equations (\ref{htilevolklmX}) and (\ref{omtilevol0lX}) can now be written as
\begin{equation}\label{VTrels}
   \frac{\partial}{\partial t} \frac{\partial^2 V^{l\nu}}{\partial x_\mu \partial x^\mu} = T^{l\nu} ,
   \qquad 
   \frac{\partial}{\partial x^l} \frac{\partial^2 V^{l\nu}}{\partial x_\mu \partial x^\mu} = - T^{0\nu} .
\end{equation}
implying that $V^{l\nu}$ and $H_{\rm t/m}$ have dimensions of mass or energy ($c=1$). As announced,  $V^{l\nu}$ is determined by the energy-momentum tensor and vanishes in the absence of matter.

Note that the three derivatives in Eq.~(\ref{VTrels}) are required to go from the level of lowest derivatives (tetrad variables) to the level of highest derivatives (conjugate tetrad variables), with the gauge vector fields and their conjugates in between (compare, for example, Eqs.~(\ref{hklevolmat}) and (\ref{htilevolklm})). Note that the different numbers of derivatives occurring in the various fields are also reflected in the different powers of $L^{-1}$ in Table~\ref{tabledimensions}.

In the presence of matter, the procedure for selecting among the solutions of the Yang-Mills theory with external fluxes extends the idea of composite theories. This selection criterion should provide stability instead of the vanishing conjugate momenta associated with the tetrad variables for the composite theory of pure gravity. Again, the selection is very restrictive so that the composite theory of gravity possesses only few degrees of freedom.

\subsection{Compact form of theory}\label{seccompactmat}
As in Section~\ref{seccompactformpure}, we would like to find a closed set of differential equations for the tetrad variables, but now in the presence of matter. Again we need to express all the Yang-Mills variables in terms of the tetrad variables. Expressions for the vector fields $A_{a \mu}$ can be obtained from the evolution equations (\ref{omklevol}), (\ref{om0kevol}), (\ref{hklevolmat}) and the primary constraints (\ref{glinconstraints1a}). Their conjugates $E^{a \mu}$ can then be extracted from the original evolution equation (\ref{Aevola0}) for the temporal components and the modified equations (\ref{Aevol0kjmod}), (\ref{Aevolkljmod}) for the spatial components of the gauge vector fields. For the convenience of the reader, the explicit representations are listed in Appendix~\ref{appYMtetrad}. By construction, these expressions satisfy the primary constraints identically.

We only need to consider the evolution equations for the conjugate Yang-Mills fields $E^a_\mu$ (the higher constraints can be verified in a straightforward manner). From Eq.~(\ref{Eevolkljmod}) we obtain
\begin{eqnarray}
   \frac{\partial}{\partial x^l}
   \left( \frac{1}{2} \frac{\partial^2 h_{jk}}{\partial x_\mu \partial x^\mu} + C_{jk} \right)
   + \delta_{jk} \frac{\partial C_{l\mu}}{\partial x_\mu} &=& \nonumber\\
   && \hspace{-12em} \frac{\partial}{\partial x^k}
   \left( \frac{1}{2} \frac{\partial^2 h_{jl}}{\partial x_\mu \partial x^\mu} + C_{jl} \right)
   + \delta_{jl} \frac{\partial C_{k\mu}}{\partial x_\mu} . \qquad 
\label{tetradsummatkl1}
\end{eqnarray}
% For the gravitational around a particle, we can go into the rest frame of the particle so that $C_{\mu\nu}$ is independent of time. The gravitational field can still depend on time, say for a gravitational wave passing a star.
By using that the tensor $T_{\mu\nu}$ in Eq.~(\ref{Ctensorchoice0}) satisfies the energy-momentum conservation (\ref{energymomconsmattT}), we obtain the following generalization of Eq.~(\ref{tetradhkl}) ,
\begin{eqnarray}
   \frac{1}{2} \frac{\partial^2 h_{kl}}{\partial x_\mu \partial x^\mu}
   + G_1 \bigg( T_{kl} - \frac{1}{2} {T^\lambda}_\lambda \, \eta_{kl} \bigg)
   + 2 G_2 {T^\lambda}_\lambda \, \eta_{kl} &=& \qquad \nonumber \\
   && \hspace{-6em} \frac{1}{2} \frac{\partial^2 f}{\partial x^k \partial x^l} ,
\label{tetradsummatkl2}
\end{eqnarray}
where the function $f$ results from integration of the third-order equations. From Eq.~(\ref{Eevolkl0}) we obtain another integrability condition,
\begin{equation}\label{tetradsummat0l1}
   \frac{\partial}{\partial x^l}
   \bigg( \frac{1}{2} \frac{\partial^2 h_{0k}}{\partial x_\mu \partial x^\mu} + G_1 T_{0k} \bigg) =
   \frac{\partial}{\partial x^k}
   \bigg( \frac{1}{2} \frac{\partial^2 h_{0l}}{\partial x_\mu \partial x^\mu} + G_1 T_{0l} \bigg) .
\end{equation}
From Eqs.~(\ref{Eevol0k0}) and (\ref{Eevol0kjmod}) we obtain after using Eqs.~(\ref{energymomconsmattT}) and (\ref{tetradhomtilrepsm1}),
\begin{eqnarray}
   \frac{\partial}{\partial x^l}
   \bigg[ \frac{1}{2} \frac{\partial^2 h_{00}}{\partial x_\mu \partial x^\mu}
   + G_1 \bigg( T_{00} - \frac{1}{2} {T^\lambda}_\lambda \, \eta_{00} \bigg) && \nonumber\\
   && \hspace{-16em} + 2 G_2 {T^\lambda}_\lambda \, \eta_{00} \bigg]
   = \frac{\partial}{\partial t}
   \left( \frac{1}{2} \frac{\partial^2 h_{0l}}{\partial x_\mu \partial x^\mu} + G_1 T_{0l} \right) ,
   \qquad 
\label{tetradsummatjl00}
\end{eqnarray}
and
\begin{eqnarray}
   \frac{\partial}{\partial t}
   \bigg[ \frac{1}{2} \frac{\partial^2 h_{jl}}{\partial x_\mu \partial x^\mu}
   + G_1 \bigg( T_{jl} - \frac{1}{2} {T^\lambda}_\lambda \, \eta_{jl} \bigg)
   + 2 G_2 {T^\lambda}_\lambda \, \eta_{jl} \bigg] &=&
   \nonumber\\
   && \hspace{-20em} \frac{\partial}{\partial x^l}
   \left( \frac{1}{2} \frac{\partial^2 h_{0j}}{\partial x_\mu \partial x^\mu} + G_1 T_{0j} \right) ,
\label{tetradsummatjl0l}
\end{eqnarray}
respectively. Again, the choice (\ref{tetradhomtilrepsm1}) of $V^{l\nu}$ is of crucial importance because it leads to further integrability conditions. Equations (\ref{tetradsummat0l1})--(\ref{tetradsummatjl0l}) allow us to extend the differential equation (\ref{tetradsummatkl2}) to all components,
\begin{eqnarray}
   \frac{1}{2} \frac{\partial^2 h_{\mu\nu}}{\partial x_\lambda \partial x^\lambda}
   + G_1 \bigg( T_{\mu\nu} - \frac{1}{2} {T^\lambda}_\lambda \, \eta_{\mu\nu}\bigg)
   + 2 G_2 {T^\lambda}_\lambda \, \eta_{kl} &=& \qquad \nonumber \\
   && \hspace{-6em} \frac{1}{2} \frac{\partial^2 f}{\partial x^\mu \partial x^\nu} ,
\label{tetradsummatall}
\end{eqnarray}
possibly after a minor modification of $f$.

The compact equation (\ref{tetradsummatall}) has a remarkable similarity with the linearized version of Einstein's field equation (\ref{linGRfieldeq2}) with the curvature tensor (\ref{linRt}) in a harmonic coordinate system, provided that we choose
\begin{equation}\label{G12choices}
   G_1 = 8 \pi G , \qquad G_2 = 0 ,
\end{equation}
and $f=0$. The freedom of choosing the function $f$ is the only leftover from the higher derivative nature of the theory. It gives us the remarkable possibility to mimic the local gauge degree of freedom associated with the general coordinate transformations employed to achieve the one-parameter family of coordinate conditions (\ref{harmoniclin}), although the composite theory is defined in Minkowski space.

\subsection{Isotropic solution revisited}
As an application of our compact equations, we consider a mass $M$ resting at the origin, which is represented by an energy-momentum tensor $T_{\mu\nu}$ with only one nonvanishing component, $T_{00} = M \delta^3(\bm{x})$. Equation~(\ref{VTrels}) requires nonzero components $V^{l0}$. A simple solution of this equation is found to be
\begin{equation}\label{isoVl0sol}
   V^{l0} = \frac{M}{8 \pi} \frac{x^l}{r} ,
\end{equation}
which describes a purely orientational effect. The complete list of conjugate tetrad variables is given by
\begin{equation}\label{isoconjtetrads}
   \tilde{h}^{0l} = \tilde{g} \tilde{\omega}^{0l} =
   - \frac{\Lambda_{\rm E} M}{4 \pi} \frac{x^l}{r^3} , \quad
   \tilde{\omega}^{kl} = \tilde{h}^{kl} = \tilde{h}^{00} = 0 .
\end{equation}
Note that the modification of the coordinate condition (\ref{h0kevolmat}) is extremely tiny, but independent of the distance from the central mass.

We now focus on the field equations (\ref{tetradsummatall}) with the parameter choices (\ref{G12choices}). Away from the origin, these equations have already been solved in Section \ref{secisostatsol}. By integrating the simplified field equations
\begin{equation}\label{tetradh00kls}
   \frac{\partial^2 h_{00}}{\partial x_n \partial x_n} + G_1 T_{00} = 0 , \quad
   \frac{\partial^2 (h_{ll}-f)}{\partial x_n \partial x_n} + 3 G_1 T_{00} = 0 ,
\end{equation}
over a sphere around the origin and using $h_{ll}-f=3(\bar{\alpha}+\bar{\xi})$, we find $r_0=MG$ and $\bar{c}=1$ for the coefficients in the solution (\ref{linbetasol}), (\ref{linalphaxisol}). More details about isotropic solutions can be found in Appendix \ref{appharmonic}.

\subsection{Modified particle motion}\label{secmodpartmot}
For obtaining the motion of a particle with mass $m$ in a gravitational field, it is convenient to divide the Hamiltonian (\ref{Hamfullsystem}) by $\Lambda_{\rm E}$ because the resulting equations then look more familiar. Whereas the variational problem of the Lagrangian approach is clearly unaffected by such a constant factor, it corresponds to a rescaling of the particle momentum variables in the Hamiltonian formulation. However, the particle trajectories remain unchanged.

We assume that the influence of the Hamiltonian $H_{\rm t/m}$ is negligibly small and only $H_{\rm m}$ and $H_{\rm YM/m}$ contribute to the particle motion. This assumption is justified by the extremely small factor $\Lambda_{\rm E}$ in $\tilde{h}^{\kappa\lambda}$ (see, for example, Eq.~(\ref{isoconjtetrads}) for static isotropic fields). The Hamiltonian $H_{\rm t/m}$ might have an influence of the motion of mass only on cosmological length and time scales.

The resulting evolution equation for the particle momentum is given by
\begin{equation}\label{fullparticleevolp}
   \frac{d p_j}{d t} = \frac{p_\mu p_\nu}{2 \gamma m} \, \frac{\partial}{\partial x^j}
   \left[ h^{\mu\nu} - \frac{2}{\Lambda_{\rm E}}
   \left( G_1 \mathring{\cal R}^{\mu\nu} + G_2 \eta^{\mu\nu} {{\cal R}^\lambda}_\lambda \right) \right] ,
\end{equation}
where we have used the expression (\ref{YML2HR}) for $H_{\rm YM/m}$, and the evolution of the particle position is governed by
\begin{eqnarray}
   \left( 1 + \frac{1}{2} h^{00} - \frac{p_k p_l h^{kl}}{2 \gamma^2 m^2} \right)
   \frac{d x^j}{d t} &=& \nonumber \\
   && \hspace{-10em} \label{fullparticleevolx}
   \left( \delta^{j\mu} -  h^{j\mu} + \frac{2 G_1}{\Lambda_{\rm E}} \mathring{\cal R}^{j \mu} \right)
   \frac{p_\mu}{\gamma m}  \\
   && \hspace{-10em}
   + \, \frac{1}{\Lambda_{\rm E}} \left[ G_1 \left( \mathring{\cal R}^{00}
   - \frac{p_k p_l \mathring{\cal R}^{kl}}{\gamma^2 m^2} \right)
   + G_2 \frac{{{\cal R}^\lambda}_\lambda}{\gamma^2} \right] 
   \frac{p_j}{\gamma m} . \nonumber
\end{eqnarray}
The factor in parentheses on the left-hand side of Eq.~(\ref{fullparticleevolx}) simply changes $d x^j/d t$ into $d x^j/d \tau$, where $\tau$ is the proper time of the particle moving in a gravitational field.

For the static isotropic solution in the weak-field approximation, the curvature tensor vanishes. Equations (\ref{fullparticleevolp}) and (\ref{fullparticleevolx}) then describe geodesic motion. However, this should not be taken for granted. For the fully nonlinear composite theory of gravity, it has been shown in Appendix~A of \cite{hco231} that only ${{\cal R}^\lambda}_\lambda$ and ${\cal R}^{00}$ vanish (however, that result was found in a standard quasi-Minkowskian coordinate system that does not satisfy the coordinate conditions (\ref{harmoniclin})). If one still wants to achieve geodesic motion then one would have to choose the scalar coupling of fields and matter through $G_2$ rather than the tensorial coupling through $G_1$. A more appealing option is to search for coordinate conditions characterizing a background Minkowski system that leads to a vanishing curvature tensor in matter-free space.

\section{Summary and conclusions}\label{secconclusions}
The main insight from this paper is this: A lot of things could go wrong with composite gravity, but they don't.

The canonical Hamiltonian formulation of the composite theory of gravity obtained by expressing the gauge vector fields of the Yang-Mills theory based on the Lorentz group in terms of tetrad or \emph{vierbein} variables requires $80$ fields, not counting any ghost fields for handling gauge conditions. A large number of constraints should arise, so that gravity has only a few degrees of freedom, but not so many that the theory would not admit any solutions. In addition to constraints associated with gauge degrees of freedom, there are constraints resulting from the composition rule. Quite miraculously, we obtain exactly the right total number of constraints. In the presence of matter, securing solutions by avoiding too many constraints requires a consistently matched double coupling of matter to both Yang-Mills and tetrad fields. The possibility of finding a proper number of natural constraints relieves the pressure to use smaller Lie groups like SU(2), which is behind the Ashtekar variables proposed for a canonical approach to gravity in the context of \emph{dreibein} variables \cite{Ashtekar86,Ashtekar87}.

Composite theories involve higher derivatives and are hence prone to instability. For composite gravity, one would expect fourth-order differential equations. However, the constraints lead to a very special feature of composite higher derivative theories: they select solutions from a workhorse theory. For composite gravity this means that we deal with selected solutions of the Yang-Mills theory based on the Lorentz group. In the presence of matter, the Yang-Mills theory includes suitable external fluxes. This selection effect guarantees the elimination of instabilities. As the selection is very restrictive, we hope that it also helps to eliminate potential problems associated with the non-compact nature of the Lorentz group (Yang-Mills theories are usually based for good reasons on compact Lie groups). As composite gravity provides selected solutions of a Yang-Mills theory, it is much closer to the standard treatment of electroweak and strong interactions than general relativity.

As a consequence of the equivalence principle, gravity is all about geometry. However, this remark does not imply that gravity must necessarily be interpreted as curvature in space-time \cite{Jimenezetal19}. The composite theory of gravity expresses the Yang-Mills fields associated with the Lorentz group in terms of the tetrad fields associated with a space-time metric. This metric is only used for expressing momenta in terms of velocities and may hence be interpreted as an anisotropy of mass. The metric has no effect on the measure used for the integrations in the Hamiltonian or Lagrangian, which are performed in an underlying Minkowski space. Nevertheless, the particle motion in the field around a central mass turns out to be geodesic. And nevertheless, the field equations for a tensorial coupling of the gravitational field to matter are remarkably similar to general relativity in the weak-field approximation. In the nonlinear regime, however, it might turn out to be necessary to use the scalar coupling to guarantee the geodesic motion of particles.

The canonical Hamiltonian formulation of the evolution equations of composite gravity in a large space is clearly advantageous for quantization. The constraints resulting from the composition rule are found to be gauge invariant, second class constraints. This suggests that, in the quantization process, they can be treated via Dirac brackets, and the gauge constraints can be treated independently with the BRST procedure. Therefore, quantization of linearized composite gravity in the context of dissipative quantum field theory \cite{hcoqft} seems to be straightforward. A compact formulation of the equations for the metric is advantageous for solving practical problems, even though these second-order differential equations have some special features: a free function appears as a result of eliminating higher derivatives by integration; this function is reminiscent of gauge degrees of freedom in general relativity.

The steps carried out here in great detail for the weak-field approximation should provide guidance for the proper canonical treatment of the fully nonlinear composite theory of gravity proposed in \cite{hco231}. Whereas many of the steps are straightforward and may actually be more transparent in the nonlinear setting (for example, true vector indices can be recognized more easily), special attention must be paid to the coordinate conditions that we want to use for characterizing appropriate Minkowskian coordinate systems (see Appendix \ref{appharmonic}). It would be desirable to find coordinate conditions for which the curvature tensor vanishes in empty space. Moreover, one needs to make a choice between coordinate conditions that are more in the spirit of general relativity or better matched to the assumption of a background Minkowski metric.

\appendix

\section{From Lagrangian to Hamiltonian for coupling of field to matter}\label{appL2H}
The goal of this appendix is to derive the contribution to the Hamiltonian that expresses the coupling of the Yang-Mills field for the Lorentz group to matter. We emanate from the following Lagrangian for a pure Yang-Mills theory,
\begin{equation}\label{LpureYM}
   L = - \int \left(  \frac{1}{4} F^a_{\mu\nu} F_a^{\mu\nu}
   + \frac{1}{2} \frac{\partial A^a_\mu}{\partial x_\mu}
   \frac{\partial A_a^\nu}{\partial x^\nu} \right) d^3x ,
\end{equation}
where, in the weak-field approximation, the field tensor is given by
\begin{equation}\label{YMfieldtensordef}
   F^a_{\mu\nu} = \frac{\partial A^a_\nu}{\partial x^\mu} -
   \frac{\partial A^a_\mu}{\partial x^\nu} .
\end{equation}
The second contribution in Eq.~(\ref{LpureYM}) represents a covariant gauge breaking term for removing degeneracies associated with gauge invariance (the particular form corresponds to the convenient Feynman gauge).

For the Yang-Mills theory based on the Lorentz group we can replace summations over $a$ by summations over $\tilde{\kappa}$, $\tilde{\lambda}$ according to Table~\ref{tabindexmatch}. If we sum over all pairs $(\tilde{\kappa}, \tilde{\lambda})$ and assume antisymmetry in $\tilde{\kappa}$, $\tilde{\lambda}$ [cf.\ Eq.~(\ref{glinA})], each term occurs twice. We include the coupling of the Yang-Mills field to matter by generalizing Eq.~(\ref{LpureYM}) to
\begin{equation}\label{LcoupledYM}
   L \! = \! - \! \int \bigg[  \frac{1}{8}
   (F^{(\tilde{\kappa}\tilde{\lambda})}_{\mu\nu}+H^{\tilde{\kappa}\tilde{\lambda}}_{\mu\nu})
   (F_{(\tilde{\kappa}\tilde{\lambda})}^{\mu\nu}+H_{\tilde{\kappa}\tilde{\lambda}}^{\mu\nu})
   + \frac{1}{2} \frac{\partial A^a_\mu}{\partial x_\mu}
   \frac{\partial A_a^\nu}{\partial x^\nu} \bigg] d^3x  ,
\end{equation}
where the fourth-rank tensor $H^{\tilde{\kappa}\tilde{\lambda}}_{\mu\nu}$ is assumed to have the same antisymmetries in $\tilde{\kappa}$, $\tilde{\lambda}$ and $\mu$, $\nu$ as $F^{(\tilde{\kappa}\tilde{\lambda})}_{\mu\nu}$. We now assume that $H^{\tilde{\kappa}\tilde{\lambda}}_{\mu\nu}$ is a linear function of the energy-momentum tensor of matter. The natural way of building a fourth-rank tensor with the required antisymmetries from a symmetric second-rank tensor $C_{\mu\nu}$ is
\begin{equation}\label{Htensorbuild}
   H^{\tilde{\kappa}\tilde{\lambda}}_{\mu\nu} = 
   {C^{\tilde{\kappa}}}_{\mu} {\delta^{\tilde{\lambda}}}_{\nu}
   - {C^{\tilde{\lambda}}}_{\mu} {\delta^{\tilde{\kappa}}}_{\nu}
   - {C^{\tilde{\kappa}}}_{\nu} {\delta^{\tilde{\lambda}}}_{\mu} 
   + {C^{\tilde{\lambda}}}_{\nu} {\delta^{\tilde{\kappa}}}_{\mu} ,
\end{equation}
where we assume that the matter tensor $C_{\mu\nu}$ is a linear combination of the trace-free and trace parts of the energy-momentum tensor $T_{\mu\nu}$. This assumption is motivated by the equations of general relativity.

Einstein's (linearized) field equation is usually written in the form
\begin{equation}\label{linGRfieldeq1}
   R_{\mu\nu} - \frac{1}{2} {R^\lambda}_\lambda \, \eta_{\mu\nu} = - 8 \pi G \, T_{\mu\nu} ,
\end{equation}
or in the alternative form
\begin{equation}\label{linGRfieldeq2}
   R_{\mu\nu}  = - 8 \pi G \,
   \left(  T_{\mu\nu} - \frac{1}{2} {T^\lambda}_\lambda \, \eta_{\mu\nu} \right) .
\end{equation}
In Eq.~(\ref{linGRfieldeq1}), the energy-momentum tensor $T_{\mu\nu}$ on the right-hand side is divergence free (conservation of energy and momentum), so that it has to be matched with the divergence free version of the curvature tensor $R_{\mu\nu}$ on the left-hand side (Bianchi identity). Equation (\ref{linGRfieldeq2}) is obtained by means of the trace equation,
\begin{equation}\label{linGRfieldeq3a}
   {R^\lambda}_\lambda = 8 \pi G \, {T^\lambda}_\lambda ,
\end{equation}
which follows by taking the trace of either version of Einstein's field equation. A particularly useful form of Einstein's field equation for our purposes is obtained by equating trace-free tensors rather than divergence-free tensors,
\begin{equation}\label{linGRfieldeq3b}
   \mathring{R}_{\mu\nu} = - 8 \pi G \, \mathring{T}_{\mu\nu} ,
\end{equation}
with the trace-free tensors
\begin{equation}\label{Rringdef}
   \mathring{R}_{\mu\nu} = R_{\mu\nu} - \frac{1}{4} {R^\lambda}_\lambda \, \eta_{\mu\nu} ,
\end{equation}
and
\begin{equation}\label{Tringdef}
   \mathring{T}_{\mu\nu} = T_{\mu\nu} - \frac{1}{4} {T^\lambda}_\lambda \, \eta_{\mu\nu} .
\end{equation}
Of course, Eq.~(\ref{linGRfieldeq3b}) now needs to be supplemented by Eq.~(\ref{linGRfieldeq3a}) to reproduce the full content of Einstein's field equations. The clear separation between trace-free and trace parts motivates our choice
\begin{equation}\label{Ctensorchoice}
   C_{\mu\nu} = G_1 \, \mathring{T}_{\mu\nu} + G_2 \, \eta_{\mu\nu}  {T^\lambda}_\lambda ,
\end{equation}
in Eq.~(\ref{Htensorbuild}), where the coefficients $G_1$, $G_2$ must have the same dimensions as Newton's constant $G$ (cf.\ Table~\ref{tabledimensions}).

From the Lagrangian (\ref{LcoupledYM}), we obtain the conjugate momenta
\begin{equation}\label{YML2conjmom}
   E^a_\nu = - \frac{\delta L}{\delta \dot{A}_a^\nu} =
   - \frac{\partial A^a_\nu}{\partial t} + \frac{\partial A^a_0}{\partial x^\nu}
   + {\delta^0}_\nu \frac{\partial A^a_\mu}{\partial x_\mu}
   - H^{\tilde{\kappa}\tilde{\lambda}}_{0\nu} .
\end{equation}
These equations can also be regarded as evolution equations for $A^a_\nu$ (due to the gauge breaking term, even for $\nu=0$). The evolution equations for $E^a_\nu$ are given by
\begin{eqnarray}\label{YML2evol}
   \frac{\partial E^a_\nu}{\partial t} &=& - \frac{\delta L}{\delta A_a^\nu} \\
   &=& \nonumber
   - \frac{\partial (F^a_{n\nu} + H^{\tilde{\kappa}\tilde{\lambda}}_{n\nu})}{\partial x_n}
   - \frac{\partial}{\partial x_\mu} \left( \frac{\partial A^a_\mu}{\partial x^\nu}
   - {\delta^0}_\nu \, \frac{\partial A^a_\mu}{\partial t} \right) .
\end{eqnarray}

We can now evaluate the Hamiltonian $H$ as the Legendre transform of $L$,
\begin{eqnarray}\label{YML2H}
   H &=& H_{\rm YM} \\ \nonumber &+& \int \!\! \left(
   \frac{1}{4} F^{(\tilde{\kappa}\tilde{\lambda})}_{mn} H^{mn}_{\tilde{\kappa}\tilde{\lambda}}
   - \frac{1}{2} E^{(\tilde{\kappa}\tilde{\lambda})}_n H^{0n}_{\tilde{\kappa}\tilde{\lambda}}
   + \frac{1}{8} H^{\tilde{\kappa}\tilde{\lambda}}_{mn} H^{mn}_{\tilde{\kappa}\tilde{\lambda}}
   \right) d^3x ,
\end{eqnarray}
where $H_{\rm YM}$ is given in Eq.~(\ref{pH2LHamf}). The contribution from
\begin{equation}\label{YML2HHH}
   \frac{1}{8} H^{\tilde{\kappa}\tilde{\lambda}}_{mn} H^{mn}_{\tilde{\kappa}\tilde{\lambda}} =
   \frac{1}{2} \big( {C^\lambda}_\nu {C^\nu}_\lambda + {C^l}_l \, {C^n}_n - {C^0}_0 \, {C^0}_0 \big) ,
   \;\; 
\end{equation}
describes a direct local self-interaction of matter. Such self-interactions are a well-known problem and should be analyzed within a careful renormalization procedure. We here simply add the corresponding contribution to the Lagrangian and thus eliminate it from the Hamiltonian. The remaining contribution characterizes the coupling between field and matter,
\begin{eqnarray}\label{YML2Hfm}
   H_{\rm YM/m} &=& \int \left(
   \frac{1}{4} F^{(\tilde{\kappa}\tilde{\lambda})}_{mn} H^{mn}_{\tilde{\kappa}\tilde{\lambda}}
   - \frac{1}{2} E^{(\tilde{\kappa}\tilde{\lambda})}_n H^{0n}_{\tilde{\kappa}\tilde{\lambda}}
   \right) d^3x \\
   &=& \int \left( F^{(\lambda n)}_{jn} {C^j}_\lambda - E^{(\lambda j)}_j {C^0}_\lambda
   - E^{(0 l)}_j {C^j}_l \right) d^3x . \quad \nonumber
\end{eqnarray}
Note that the fields $F^{(\lambda n)}_{jn}$ contain only spatial derivatives of the spatial components of $A^a_\mu$, and no time derivatives.

This expression for $H_{\rm YM/m}$ suggests that
\begin{equation}\label{Rmunuexpress1}
   {R^\mu}_\nu = F^{(\mu\lambda)}_{\nu\lambda} ,
\end{equation}
is an interesting tensor to look at.
By using Eqs.~(\ref{glinA}) and (\ref{YMfieldtensordef}), we arrive at the following explicit representation in terms of the metric,
\begin{equation}\label{linRt}
   {R^\mu}_\nu = \frac{1}{2} \left(
   \frac{\partial^2 {h^\mu}_\nu}{\partial x_\lambda \partial x^\lambda}
   - \frac{\partial^2 {h^\mu}_\lambda}{\partial x_\lambda \partial x^\nu}
   - \frac{\partial^2 {h^\lambda}_\nu}{\partial x_\mu \partial x^\lambda}
   + \frac{\partial^2 {h^\lambda}_\lambda}{\partial x_\mu \partial x^\nu}
   \right) ,
\end{equation}
which can be recognized as the Ricci curvature tensor in the weak-field approximation [see, e.g., Eq.~(7.6.2) of \cite{Weinberg} or Eq.~(B5) of \cite{hco231}; cf.\ also Eq.~(\ref{GRgeneq})]. By means of Eqs.~(\ref{Rmunuexpress1}) and (\ref{YML2conjmom}), we can evaluate
\begin{eqnarray}
   {R^\mu}_\nu {C^\nu}_\mu &=& F^{(\lambda n)}_{jn} {C^j}_\lambda
   - E^{(\lambda j)}_j {C^0}_\lambda - E^{(0 l)}_j {C^j}_l \nonumber \\
   &-& ( H^{0\mu}_{0\nu} + H^{\mu j}_{0j} \, {\delta^0}_\nu ) \, {C^\nu}_\mu .
\label{RRmotivation}
\end{eqnarray}
If we define
\begin{equation}\label{RRrelation}
   \mbox{${\cal R}^\mu$}_\nu = {R^\mu}_\nu +  H^{0\mu}_{0\nu} + H^{\mu j}_{0j} \, {\delta^0}_\nu ,
\end{equation}
the Hamiltonian (\ref{YML2Hfm}) can be rewritten as
\begin{equation}\label{YML2HR}
   H_{\rm YM/m} = \int \mbox{${\cal R}^\mu$}_\nu {C^\nu}_\mu \, d^3x .
\end{equation}
For pure gravity, that is, in the absence of matter or for $H^{\tilde{\kappa}\tilde{\lambda}}_{\mu\nu}=0$, the tensor $\mbox{${\cal R}^\mu$}_\nu$ coincides with the curvature tensor ${R^\mu}_\nu$. This simple direct coupling of curvature tensor and energy-momentum tensor suggests that the developments of this appendix are very natural and appealing.

Note that the arguments given in this appendix are not restricted to the weak-field approximation (\ref{YMfieldtensordef}). Generalization to the full theory is straightforward. Even the formula (\ref{Rmunuexpress1}) can be generalized (for $\tilde{g}=1$).

\section{Representation of Yang-Mills fields in terms of tetrad variables}\label{appYMtetrad}
In addition to the modified composition rule for the gauge vector fields resulting from the presence of matter,
\begin{equation}\label{representA0lmu}
   A_{(0l) \mu} = \frac{1}{2} \left( \frac{\partial h_{l\mu}}{\partial t} -
   \frac{\partial h_{0\mu}}{\partial x^l} \right)
   + \frac{1}{2\tilde{g}} \, \frac{\partial \omega_{0l}}{\partial x^\mu} - \Lambda_{\rm E} V_{l\mu} ,
\end{equation}
\begin{equation}\label{representAklmu}
   A_{(kl) \mu} = \frac{1}{2} \left( \frac{\partial h_{l\mu}}{\partial x^k} -
   \frac{\partial h_{k\mu}}{\partial x^l} \right)
   + \frac{1}{2\tilde{g}} \, \frac{\partial \omega_{kl}}{\partial x^\mu} ,
\end{equation}
we have the representation of their conjugate momenta obtained from the evolution equations (\ref{Aevola0}), (\ref{Aevol0kjmod}) and (\ref{Aevolkljmod}):
\begin{equation}\label{representE0l0}
   E_{(0l) 0} = \frac{1}{2} \frac{\partial}{\partial x_\mu}
   \left( \frac{\partial h_{l\mu}}{\partial t}
   - \frac{\partial h_{0\mu}}{\partial x^l} \right)
   - \Lambda_{\rm E} \frac{\partial V_{l\mu}}{\partial x_\mu} ,
\end{equation}
\begin{equation}\label{representEkl0}
   E_{(kl) 0} = \frac{1}{2} \frac{\partial}{\partial x_\mu}
   \left( \frac{\partial h_{l\mu}}{\partial x^k}
   - \frac{\partial h_{k\mu}}{\partial x^l} \right) ,
\end{equation}
\begin{eqnarray}
   E_{(0l) j} &=& \frac{1}{2} \left( \frac{\partial^2 h_{0j}}{\partial x^l \partial t}
   - \frac{\partial^2 h_{lj}}{\partial t^2}
   - \frac{\partial^2 h_{00}}{\partial x^j \partial x^l}
   + \frac{\partial^2 h_{0l}}{\partial x^j \partial t} \right) \nonumber \\
   &+& C_{jl} - \delta_{jl} \, C_{00}
   + \Lambda_{\rm E} \left( \frac{\partial V_{lj}}{\partial t}
   - \frac{\partial V_{l0}}{\partial x^j} \right) ,
\label{representE0lj}
\end{eqnarray}
and
\begin{eqnarray}
   E_{(kl) j} &=& \frac{1}{2} \left( \frac{\partial^2 h_{kj}}{\partial x^l \partial t}
   - \frac{\partial^2 h_{lj}}{\partial x^k \partial t}
   - \frac{\partial^2 h_{0k}}{\partial x^j \partial x^l}
   + \frac{\partial^2 h_{0l}}{\partial x^j \partial x^k} \right) \nonumber \\
   &+& \delta_{jk} \, C_{0l} -  \delta_{jl} \, C_{0k} .
\label{representEklj}
\end{eqnarray}

\section{Static isotropic solution in harmonic coordinates}\label{appharmonic}
Static isotropic solutions play an important role in the theory of gravity. They are the starting point (i) for many of the predictions that have been tested with high precision and (ii) for the theory of black holes. We here offer a few remarks on the role of coordinate conditions in the fully nonlinear composite theory of gravity for static isotropic solutions.

We start with the static isotropic solutions of the Yang-Mills theory, from which the solutions of the composite theory are then selected. We assume that these solutions are of the form
\begin{equation}\label{Aisotropiceleg}
   A^a_\nu = Y(r) \, T^a_{l\nu} \, x^l ,
\end{equation}
with $r=(x_1^2+x_2^2+x_3^2)^{1/2}$. More explicitly, by means of the definition (\ref{Lorentzgenerators}), the $24$ components of $A^a_\nu$ can be listed in matrix form,
\begin{equation}\label{Aisotropicelegmatr}
   A^a_\nu = Y \left(  \begin{matrix}
            x_1 & x_2 & x_3 & 0 & 0 & 0 \\
            0 & 0 & 0 & 0 & -x_3 & x_2 \\
            0 & 0 & 0 & x_3 & 0 & -x_1 \\
            0 & 0 & 0 & -x_2 & x_1 & 0 \\
            \end{matrix} \right) ,
\end{equation}
where the index $a$ of the Lie algebra (see Table~\ref{tabindexmatch}) labels the columns and the space-time index $\nu$ labels the rows. The function $Y$ has to be determined from the Yang-Mills equations for the gauge vector field (\ref{Aisotropiceleg}). The field equations for gauge vector fields of this form and their solutions have been discussed in Section~V of \cite{hco231}. Most remarkable is the closed-form solution of the fully nonlinear equations,
\begin{equation}\label{YMexactsolYZ}
   Y = \frac{1}{r^2( \tilde{g} +r/r_0 )} ,
\end{equation}
with a free parameter $r_0$, which is closely related to the Schwarzschild radius.

As a next step, one should choose the form of the static isotropic metric to be used in the composition rule for the gauge vector field (\ref{Aisotropiceleg}). It is well known from general relativity that the proper form of the isotropic metric depends on the choice of coordinates (see, e.g., Eq.~(8.1.3) of \cite{Weinberg}). The choice of coordinates does not matter in general relativity, where general coordinate transformations are possible, but is does matter in the composite theory of gravity, where only Lorentz transformations in the background Minkowski space are allowed. We here compare standard quasi-Minkowskian coordinates (see Section 8.1 of \cite{Weinberg}) and harmonic coordinates.

In the previous work \cite{hco231}, we used standard quasi-Minkowskian coordinates for associating a metric with the Yang-Mills solution (\ref{Aisotropicelegmatr}), (\ref{YMexactsolYZ}). An important conclusion was that $\tilde{g}$ should approach $0$ to reproduce the high-precision predictions of general relativity and that particularly nice black hole solutions result when $0$ is approached from below. However, these conclusions depend on the assumption that composite gravity can be applied meaningfully in standard quasi-Minkowskian coordinates (as their name might suggest).

For comparison, we here consider harmonic coordinates, which can be defined in more general situations and may be regarded as nearly Minkowskian (see, e.g., pp.\,163 and 254 of \cite{Weinberg}). We assume the following form of a static isotropic metric (see, e.g., Eq.~(8.1.3) of \cite{Weinberg}), which is sufficiently general for imposing harmonic coordinate conditions:
\begin{equation}\label{isoxg}
   g_{\mu\nu} = \left( \begin{matrix}
   -\beta & 0 \\
   0 & \alpha \, \delta_{mn} + \xi \, \frac{x_m x_n}{r^2}
   \end{matrix} \right) ,
\end{equation}
with inverse
\begin{equation}\label{isoxginv}
   \bar{g}^{\mu\nu} = \left( \begin{matrix}
   -\frac{1}{\beta} & 0 \\
   0 & \frac{\delta_{mn}}{\alpha} - \frac{\xi}{\alpha(\alpha+\xi)} \, \frac{x_m x_n}{r^2}
   \end{matrix} \right) .
\end{equation}
In our previous work based on standard quasi-Minkowskian coordinates, we assumed $\alpha=1$, $\beta=B$, and $\xi=A-1$ \cite{hco231}. We are now interested in solutions $g_{\mu\nu}$ of the nonlinear theory that satisfy the harmonic coordinate conditions
\begin{equation}\label{harmonic}
   \bar{g}^{\mu\nu} \frac{\partial g_{\rho\nu}}{\partial x^\mu} =
   \frac{1}{2} \bar{g}^{\mu\nu} \frac{\partial g_{\mu\nu}}{\partial x^\rho} .
\end{equation}
According to these conditions, the functions in Eq.~(\ref{isoxg}) are related by the differential equation
\begin{equation}\label{harmoniciso}
   \frac{\xi'}{\alpha+\xi} - \frac{\alpha+2\xi}{\alpha+\xi} \, \frac{\alpha'}{\alpha}
   + \frac{4\xi}{r\alpha} = \frac{\beta'}{\beta} ,
\end{equation}
where a prime on a function of $r$ indicates the derivative with respect to $r$.

The Schwarzschild solution of general relativity in harmonic coordinates is given by (see, e.g., Eq.~(8.2.15) of \cite{Weinberg})
\begin{equation}\label{Schwarzschildiso}
   \alpha = \left( 1 + \frac{r_0}{r} \right)^2 , \quad 
   \beta = \frac{r-r_0}{r+r_0} , \quad
   \xi = \frac{r+r_0}{r-r_0} \, \frac{r_0^2}{r^2} .
\end{equation}
One can easily verify that the functions given in Eq.~(\ref{Schwarzschildiso}) indeed satisfy Eq.~(\ref{harmoniciso}). For the Schwarzschild solution, we moreover have $(\alpha+\xi)\beta=1$.

Note that the harmonic coordinate conditions (\ref{harmonic}) are Lorentz covariant. The same would be true for the following class of simpler coordinate conditions,
\begin{equation}\label{harmonicalt}
   \eta^{\mu\nu} \frac{\partial g_{\rho\nu}}{\partial x^\mu} =
   K \eta^{\mu\nu} \frac{\partial g_{\mu\nu}}{\partial x^\rho} ,
\end{equation}
% https://en.wikipedia.org/wiki/Coordinate_conditions
which leads to
\begin{equation}\label{harmonicaltiso}
   (3K-1) \alpha'+ K \beta' + (K-1) \xi' = \frac{2 \xi}{r} .
\end{equation}
As pointed out before, in composite gravity, solutions for different coordinate conditions are not equivalent. The condition (\ref{harmonic}) is very much inspired by the thinking of general relativity. Once the decision in favor of a background Minkowski space has been made, Eq.~(\ref{harmonicalt}) may actually be the more appropriate choice.

Our construction of Yang-Mills fields is based on the symmetric tetrad variables obtained by factorizing the metric (\ref{isoxg}),
\begin{equation}\label{isoxb}
   {b^\kappa}_\mu = \left( \begin{matrix}
   \sqrt{\beta} & 0 \\
   0 & \sqrt{\alpha} \, \delta_{km} + (\sqrt{\alpha+\xi} - \sqrt{\alpha} ) \frac{x_k x_m}{r^2}
   \end{matrix} \right) ,
\end{equation}
with inverse
\begin{equation}\label{isoxbinv}
   \mbox{$\bar{b}^\mu$}_\kappa = \left( \begin{matrix}
   \frac{1}{\sqrt{\beta}} & 0 \\
   0 & \frac{\delta_{mk}}{\sqrt{\alpha}} + (\frac{1}{\sqrt{\alpha+\xi}} - \frac{1}{\sqrt{\alpha}} )
   \frac{x_m x_k}{r^2}
   \end{matrix} \right) .
\end{equation}
The nonlinear decomposition rule
\begin{eqnarray}
   A_{a \nu} {T^a}_{\kappa\lambda} &=& \frac{1}{2} \,
   \mbox{$\bar{b}^\mu$}_\kappa \left( \frac{\partial g_{\mu\nu}}{\partial x^{\mu'}}
   -\frac{\partial g_{\mu'\nu}}{\partial x^\mu} \right) \mbox{$\bar{b}^{\mu'}$}_\lambda
   \nonumber\\
   &+& \frac{1}{2 \tilde{g}} \, \frac{\partial {b^{\kappa'}}_\mu}{\partial x^\nu}
   \left( \mbox{$\bar{b}^\mu$}_\kappa \, \eta_{\kappa'\lambda}
   - \mbox{$\bar{b}^\mu$}_\lambda \, \eta_{\kappa'\kappa} \right) , \qquad 
\label{Avecdef}
\end{eqnarray}
leads to two equivalent representations of $Y$,
\begin{equation}\label{Yexpression}
   Y = \frac{1}{2 r^2} \frac{\xi-r\alpha'}{\sqrt{(\alpha+\xi)\alpha}} +
   \frac{1}{\tilde{g} \, r^2} \bigg( 1 - \frac{1}{2} \sqrt{\frac{\alpha}{\alpha+\xi}}
   - \frac{1}{2} \sqrt{\frac{\alpha+\xi}{\alpha}} \bigg) ,
\end{equation}
and
\begin{equation}\label{Zexpression}
   Y = \frac{1}{2 r} \frac{\beta'}{\sqrt{(\alpha+\xi)\beta}} .
\end{equation}

For harmonic coordinates, we find the second-order Robertson expansions
\begin{equation}\label{harmexpansions}
   \alpha = 1 + \frac{2r_0}{r} - \tilde{g} \frac{r_0^2}{r^2} , \quad 
   \beta  = 1 - \frac{2r_0}{r} + \tilde{g} \frac{r_0^2}{r^2} , \quad 
   \xi = \frac{4r_0^2}{r^2} .
\end{equation}
Similar expansions can be obtained for the coordinate conditions (\ref{harmonicalt}), provided that $K=1/2$. By matching the terms that contribute to the high-precision predictions of general relativity with the expansions of the Schwarzschild solution (\ref{Schwarzschildiso}), we find $\tilde{g}=2$ from the second-order expansion of $\beta$.

% \bibliography{hcopubs}

\end{document}